\documentclass[prb,10pt,notitlepage,twocolumn,superscriptaddress]{revtex4}

\usepackage{graphicx}
\usepackage{amsmath}
\usepackage{amssymb}
\usepackage{amsfonts}
\usepackage{color}
\usepackage{xspace}
\usepackage{subfigure}
\usepackage{bm}
\usepackage{bbold}
\usepackage{blkarray}
\usepackage{pdflscape}

\newcommand{\tJ}{$t$-$J$\xspace}
\newcommand{\bra}[1]{\langle #1 |}
\newcommand{\ket}[1]{| #1 \rangle}

\newcommand{\Mo}{Mo$_3$S$_7$(dmit)$_3$\xspace}

\newcommand{\DM}{DM\xspace}

\renewcommand{\cite}[1]{[\onlinecite{#1}]}

\begin{document}

\title{Heisenberg and Dzyaloshinskii-Moriya interactions controlled by molecular packing in tri-nuclear organometallic clusters}

\author{B. J. Powell}\affiliation{School of Mathematics and Physics, The University of Queensland, Brisbane, Queensland, 4072, Australia}

\author{J. Merino}
\affiliation{Departamento de F\'isica Te\'orica de la Materia Condensada, Condensed Matter Physics Center (IFIMAC) and
	Instituto Nicol\'as Cabrera, Universidad Aut\'onoma de Madrid, Madrid 28049, Spain}

\author{A. L. Khosla}\affiliation{School of Mathematics and Physics, The University of Queensland, Brisbane, Queensland, 4072, Australia}

\author{A. C. Jacko}\affiliation{School of Mathematics and Physics, The University of Queensland, Brisbane, Queensland, 4072, Australia}

\begin{abstract}
Motivated by recent synthetic and theoretical progress we consider magnetism in crystals of multi-nuclear organometallic complexes. We calculate the Heisenberg symmetric exchange and the Dzyaloshinskii-Moriya antisymmetric exchange. We show how, in the absence of spin-orbit coupling, the interplay of electronic correlations and quantum interference leads to a quasi-one dimensional effective spin model in a typical tri-nuclear complex, \Mo, despite its underlying three dimensional band structure.
 We show that both intra- and inter-molecular spin-orbit coupling can cause an effective Dzyaloshinskii-Moriya interaction. Furthermore, we show that, even for an isolated pair of molecules the relative orientation of the molecules controls the nature of the Dzyaloshinskii-Moriya coupling.  We show that  interference effects also play    a crucial role in determining the Dzyaloshinskii-Moriya interaction. Thus, we argue, that multi-nuclear organometallic complexes represent an ideal platform to investigate the effects of Dzyaloshinskii-Moriya interactions on quantum magnets.
\end{abstract}

\maketitle

\section{Introduction}

There has recently been significant interest in the Dzyaloshinskii-Moriya (DM) interaction  \cite{Moriya} in organic magnets  \cite{WinterReview,Winter12,Winter15,SlicterPRL,SlicterPRB,Kagawa}. At first sight it may be  surprising that the DM interaction, an  effect due to spin-orbit coupling (SOC), is significant in organic materials as they  typically contain only light elements, in which SOC is weak. However, one should note that in organic materials all energy scales are typically much smaller than in atomic crystals  \cite{PowellJPCM}. For example, organic charge transfer salts show very similar physics to the cuprate high-temperature superconductors, but with all energy scales [Ne\'el temperature, superconducting critical temperature, upper critical field, nearest neighbor hopping integral ($t$), effective on-site Coulomb interaction ($U$), etc.]  an order of magnitude smaller in the organics  \cite{McKenzieScience}. 
It is therefore interesting to ask what role SOC plays in organo-metallic magnets where the small energy scales due to the small $t$ and $U$ typical of molecular systems may be combined with the larger SOC associated with metals.

Multinuclear coordination complexes (i.e., coordination complexes containing multiple transition metals) have the potential to realize a wide range of exotic many-body physics \cite{Amie16,Jaime16}. Until recently the primary focus has been on single molecule magnetism  \cite{Christou,Sessolia}, but an emerging paradigm is the fabrication of multi-nuclear clusters with ligands that facilitate intermolecular charge transport  \cite{Llusar,LlusarReview,Almeida}. 
Intermolecular hopping integrals are typically rather small ($\lesssim0.1$~eV)  \cite{JackoPRB15} which means that, even if the {absolute} values of the parameters describing intramolecular interactions {are smaller than those for inorganic materials}, electronic correlations will be strong, cf. organic charge transfer salts  \cite{RPP}. 
For example, density functional calculations predict that \Mo is metallic  \cite{Llusar,JackoPRB15} - but, experimentally, it is found to be an insulator with a charge gap $\sim150$~K. 

The DM antisymmetric exchange interaction, $H_{DM}={\bm D}_{ij}\cdot{\bm S}_i\times{\bm S}_j$, results from the exchange of angular momentum between the spin and orbital degrees of freedom of a system \cite{Moriya}. This interaction favors the alignment of the spins ${\bm S}_i$ and ${\bm S}_j$ perpendicular to one another; and mutually perpendicular to $\bm D$, inducing an easy plane anisotropy. Therefore, the DM interaction can have important consequences, in both the ferromagnetic and antiferromagnetic cases, even when the usual (Heisenberg) symmetric exchange is {significantly} large{r}. For example, it has been argued that even moderate DM interactions ($|{\bm D}|$ a few percent of $J$) can drive long-range antiferromagnetic ordering in some frustrated antiferromagnets  \cite{Cepas,Trumper}.

A long-standing problem for strongly correlated molecular materials is how to determine the relevant simple model that captures the essential physics of the material. The construction of Wannier orbitals has proven a powerful tool as it allows one to derive tight-binding models from first-principles calculations without having to guess what the appropriate model is or needing to fit parameters to a guessed model   \cite{JackoReview}. 

\begin{figure}
	\begin{center}
		\includegraphics[width=0.9\columnwidth]{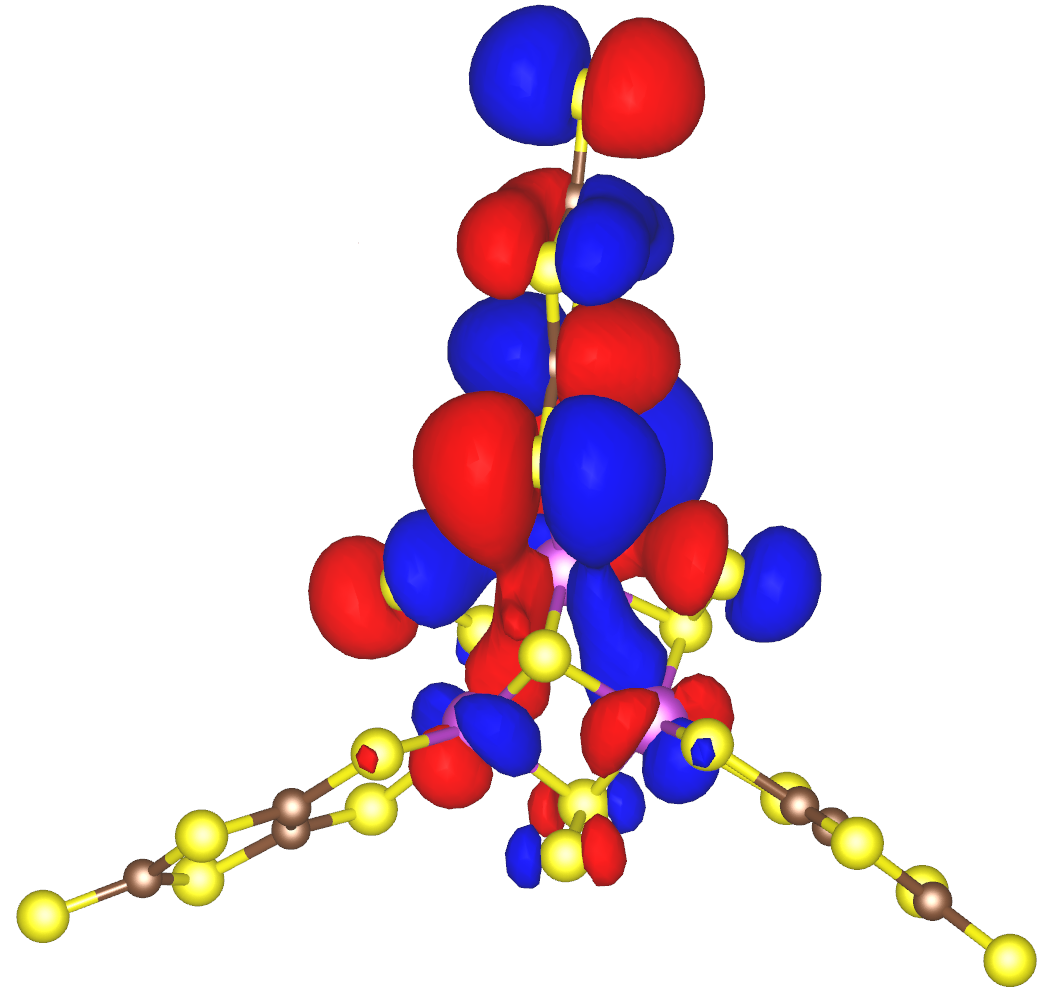}
	\end{center}
	\caption{A Wannier orbital on a single \Mo complex. The other two Wannier orbitals are related by the $C_3$ rotational symmetry of the molecule. Data from Ref. \onlinecite{Jacko16}.} 
	\label{fig:Wannier}
\end{figure}

\begin{figure}
	\begin{center}
		\includegraphics[width=0.8\columnwidth]{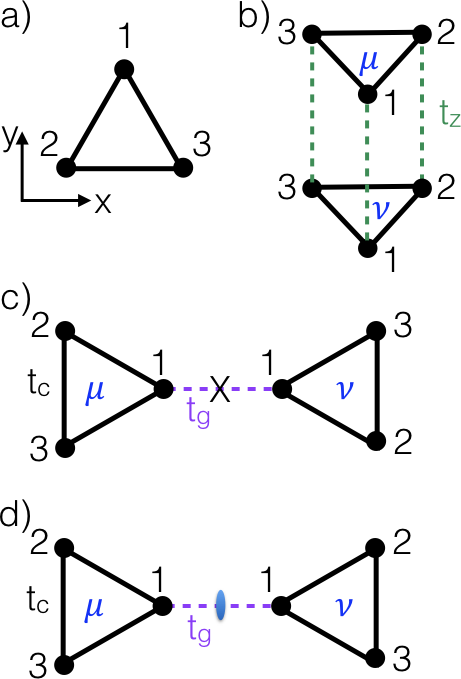}
	\end{center}
	\caption{Sketches of (the tight-binding terms in) the effective Hamiltonians discussed in this paper. (a) A single molecule. The local  $x$ and $y$ axes, defined by the phase convention for the SMOC [Eq. (\ref{HSO})], are shown. (b) Interlayer coupling model discussed in section \ref{sect:interlayer}, see particularly Eq. (\ref{HamZ}). {(c) Inversion symmetric interlayer coupling discussed in section \ref{sect:inplane}, cf. Eq. (\ref{HamG}). The inversion center is marked by the X. (d) $C_2^z$ symmetric interlayer coupling discussed in section \ref{sect:rot}. The rotation axis is marked by the oval. Note that in panel (c) the numbering on both molecules runs counterclockwise, whereas in panel (d) the numbering is clockwise on the molecule labeled $\nu$.}
	}
	\label{fig:lattice}
\end{figure}

\begin{figure}
	\begin{center}
		\includegraphics[height=0.4\columnwidth]{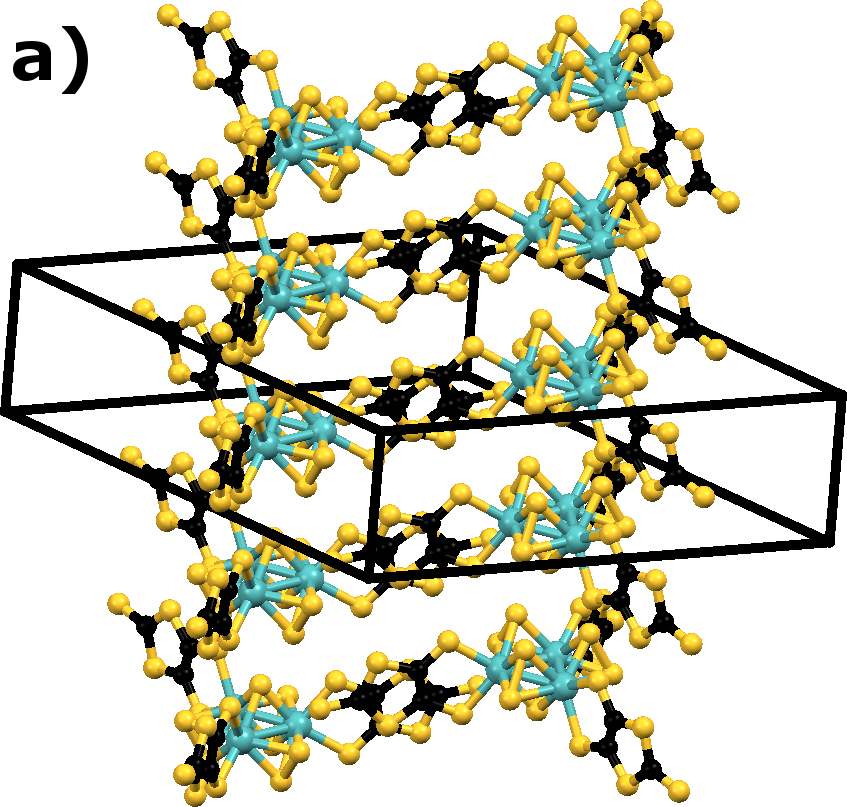}
		\includegraphics[height=0.4\columnwidth]{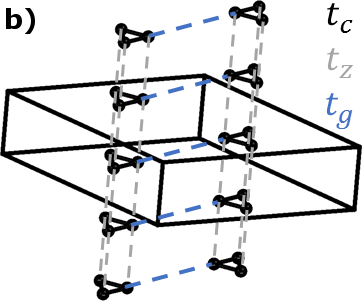}\\\vspace*{10pt}
		\includegraphics[width=0.9\columnwidth]{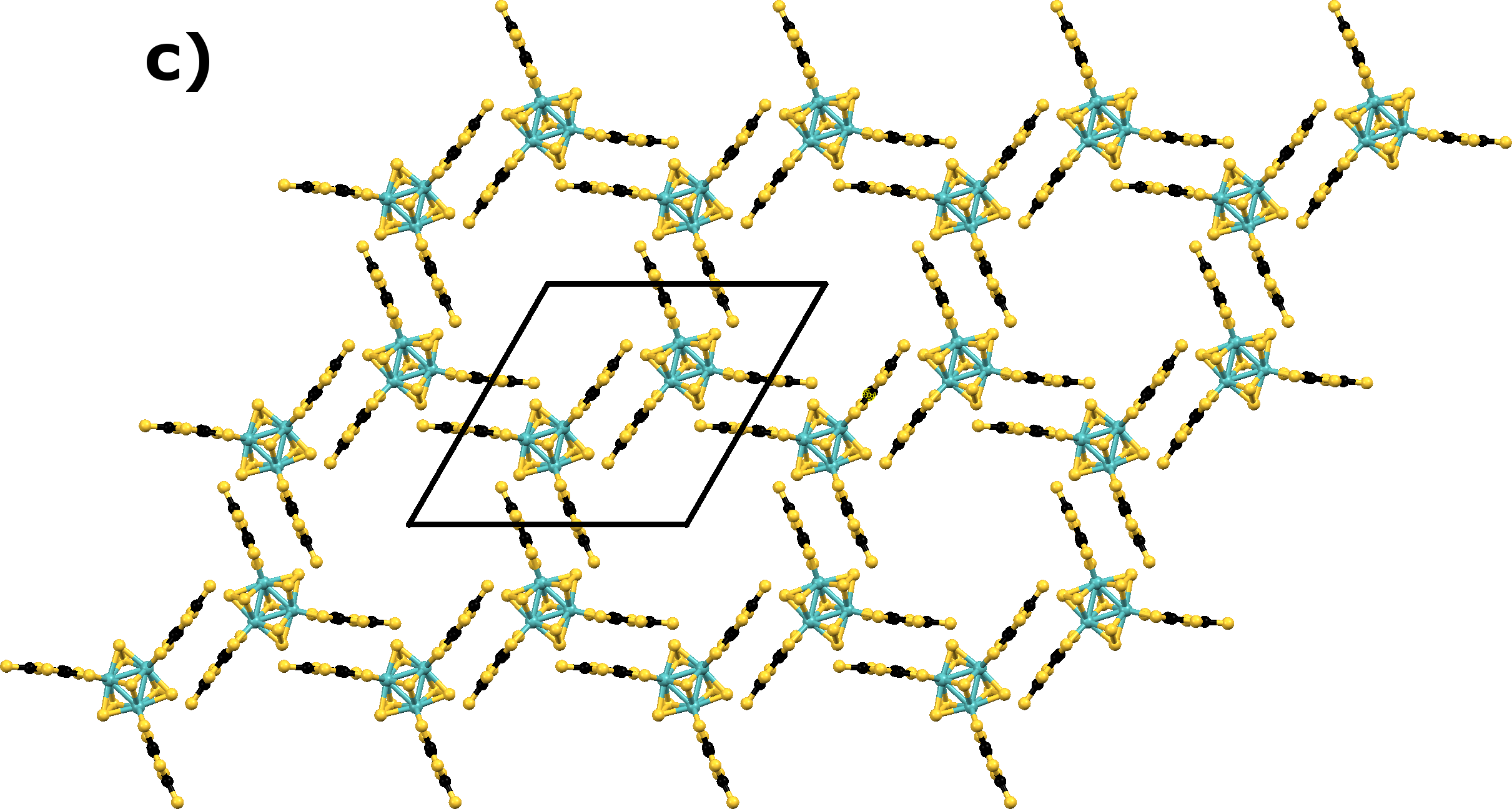} \\\vspace*{10pt}
		\includegraphics[width=0.75\columnwidth]{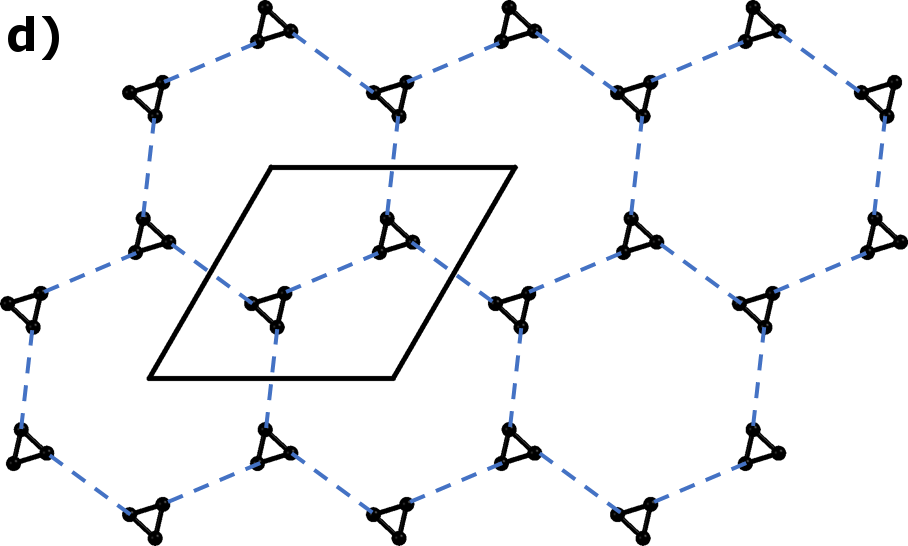}
	\end{center}
	\caption{{Crystal structure of \Mo, after \cite{Llusar}. (a,b) Triangular tubes perpendicular to the plane (along $c$-axis). (c,d) Truncated hexagonal net within a plane.  (a,c) The full molecular structure: with carbons black, sulphurs yellow and molybdenums cyan. (b,d) Simplified model showing only the Mo atoms, which are in a one-to-one correspondence to the Wannier orbitals, cf. Fig. \ref{fig:Wannier}. A unit cell is marked in all panels.
	}}
	\label{fig:xtal}
\end{figure}

A tight-binding model for \Mo derived from the Wannier orbitals calculated from density functional theory (DFT) has recently been reported  \cite{JackoPRB15}. This model contains three molecular orbitals per \Mo molecule (cf. Figs. \ref{fig:Wannier} and \ref{fig:lattice}a and Table \ref{tab:Jacko}).
{\Mo forms layered crystals, Fig. \ref{fig:xtal}. Within the $ab$-plane Mo molecules form a corrugated hexagonal lattice, with an inversion center between nearest neighbors, Fig. \ref{fig:xtal}c,d. In contrast  along the $c$-axis molecules are related to one another by translational symmetry, Fig. \ref{fig:xtal}a,b. This leads to rather different intra- and inter-layer electronic molecular couplings. The in-plane hopping, 
 $t_g$  links single vertices on neighboring triangles (Fig. \ref{fig:lattice}c), which leads to a decorated hexagonal lattice in-plane, Fig. \ref{fig:xtal}d; whereas the interlayer hopping $t_z$ connects each Wannier orbital with the equivalent orbital in the unit cell above it (Fig. \ref{fig:lattice}b), leading to triangular tubes of Wannier orbitals perpendicular to the plain (Fig. \ref{fig:xtal}b).}

\begin{table}
	\centering
	\begin{tabular}{ccc} 
		\hline\hline
		parameter  & \hspace{1cm} &	value [meV]   \\ \hline 
	     $t_c$  && 60   \\
	     $t_g$ && 47     \\
	     $t_z$     & &  41    \\
	     $\lambda_z$ &&  4.9    \\
	     $\lambda_{xy}$     & & 2.5     \\
		\hline\hline
	\end{tabular}
	\caption{{Parameters of the minimal tight-binding model for \Mo, from \cite{JackoPRB15,Jacko16}. This model is sketched in Figs. \ref{fig:lattice} and \ref{fig:xtal}b,c and defined by Eqs. (\ref{HtJ}), (\ref{HamG}), and (\ref{HamZ}).}
	}
	\label{tab:Jacko}
\end{table}

On average, four electrons occupy the three Wannier orbitals per cluster -- therefore, DFT predicts that \Mo is metallic in the absence of long-range antiferromagnetic order, which is not observed experimentally. On the other hand, \Mo is found to be an insulator experimentally  \cite{Llusar}.
However, \Mo cannot be a {simple} Mott insulator as the relevant tight-binding model is two-thirds filled, i.e., four electrons per triangular molecule  \cite{JackoPRB15}. 
The three site Hubbard model with four electrons has a triplet ground state for any $t>0$ and $U>0$  \cite{MerinoPRB06}.
The Wannier/DFT calculations suggest that, in \Mo, hopping in the plane ($t_g$) is similar to (indeed slightly larger than) the interlayer hopping ($t_z$){, cf. Table \ref{tab:Jacko}}. Surprisingly, we have recently found that correlations drive \Mo to the quasi-one dimensional limit in the insulating phase  \cite{Jaime16}. Our analytical treatment, below, will allow for a full elucidation of the mechanism behind this.  
{In particular, we will demonstrate that the emergent one-dimensionality of \Mo is a consequence of the interplay between the internal molecular electronic structure and its crystal structure.}

Furthermore, it has recently been shown that in one dimension  the tendency towards molecular triplet formation remains and drives an insulating state with spin-one molecular moments  \cite{JananiPRL,JananiPRB,Henry}. In particular the model is insulating at two-thirds filling even when only \emph{on-site} interactions are included -- no charge order is predicted  \cite{JananiPRL,JananiPRB,Henry}. In the insulating phase the molecular moments are  coupled by an effective Heisenberg interaction  \cite{JananiPRB} causing the one-dimensional model to realize the Haldane phase  \cite{JananiPRL}.

In order to calculate the DM interaction we must first understand how SOC enters the problem. One possibility is via intra-atomic effects on the Mo or perhaps the S atoms; but these are suppressed by the large energy gaps to the atomic excited states. A more interesting route is that spin couples directly to the angular momenta associated with the currents running around the three Wannier orbitals on each molecule  \cite{Amie16}. Indeed, it has  been shown that this  spin molecular-orbital coupling (SMOC)  provides an accurate description of the (single particle) electronic structure of \Mo  \cite{Jacko16,Amie16}.

Given the flexibility of organometallic chemistry, it is natural to ask what physics could be relevant in other tri-nuclear complexes. For example, the selenated analogues of \Mo have been synthesized, but little is known about their magnetic properties. Further, it has been shown from first principles that the SMOC is greatly enhanced in tungsten tri-nuclear complexes  \cite{Amie16}.

In this paper we consider the general problem of two neighboring tri-nuclear complexes at two-thirds filling. We consider both in-plane (Fig. \ref{fig:lattice}c{, d}) and inter-layer  (Fig. \ref{fig:lattice}b) coupling. We show that  the combination of electron-electron interactions, SOC and intermolecular hopping can lead to a DM interaction between neighboring spin-one molecular moments (when not symmetry forbidden). In this sense the molecules act like complex artificial atoms. In natural atoms, the relative orientation of atoms is defined by the orientation of their local environments, e.g., the oxygen octahedra in iridates, cf.  \cite{Perkins}. In organic systems the relative orientation of the molecules drives large changes in the DM interaction between molecules, independent of their local environments. Furthermore, the inherent flexibility of the molecular platform suggests that synthetic chemistry will allow one to tune the interactions so as to enhance or suppress {particular} physical effects. 

The remainder of this paper is laid out as follows: In section \ref{sect:1mol} we study the \tJ model of a single molecule in the presence of SMOC. In section \ref{sect:2mol-SMOC}, we consider the effects of SMOC on the effective interactions between a pair of tri-nuclear clusters. Here we show that the packing motif of the crystal has a dramatic effect on the \DM interactions between neighboring clusters. We also explain how the interplay of electronic correlations and quantum mechanical interference leads to a quasi-one dimensional effective Hamiltonian, even if the material has an underlying three dimensional electronic structure.  In section \ref{sect:2mol-SOI} we allow for a fully general SOC and show that intermolecular SOC can also lead to \DM interactions.

\section{Single molecules}\label{sect:1mol}

As {discussed} above, an accurate tight-binding model of \Mo can be constructed with just three Wannier orbitals per molecule. The Wannier orbitals are hybrids of the dmit molecular orbitals with a single d-orbital per Mo and  orbitals on the S atoms in the core (see Fig. \ref{fig:Wannier}). It is  natural that only a single d-orbital per Mo atom contributes to the Wannier orbitals as the Mo atoms sit in low symmetry environments and so the degeneracy of the Mo d-orbitals is entirely lifted. These arguments apply to many multi-nuclear complexes; therefore, one expects that a three orbital description will suffice in many tri-nulcear complexes  \cite{Hoffman}.

Thus, we model the $\mu$th molecule by the three-site single band electron doped \tJ model  \cite{Essler}, cf. Fig. \ref{fig:lattice}a:
\begin{widetext}
\begin{eqnarray}
H_{tJ}^{(\mu)}&\equiv& P_0\left[\sum_{\sigma,j=1}^{3}t_c\left(\hat{a}^\dagger_{\mu j\sigma}\hat{a}_{\mu j+1\sigma}+\hat{a}^\dagger_{\mu j\sigma}\hat{a}_{\mu j-1\sigma}\right) 
%\right. \notag\\&& \hspace*{1cm}
+
\frac{J_c}{4}\sum\limits_{i\ne j\ne k=1}^{3}\sum_{\sigma,\sigma'} \hat{a}_{\mu i\sigma} \hat{a}^\dagger_{\mu j\sigma} (1-\hat n^{(a)}_{\mu j\uparrow})
% \notag\\&& \left. \hspace*{2cm}
% \times
(1-\hat n^{(a)}_{\mu j\downarrow})\hat a_{\mu j\sigma'}\hat a^\dagger_{\mu k\sigma'}
\right]P_0
\label{HtJ}
\end{eqnarray}
%\begin{widetext}
%\begin{eqnarray}
%H_{tJ}&\equiv& P_0\left[\sum_{\sigma,j=1}^{3}t_c\left(\hat{a}^\dagger_{\mu j\sigma}\hat{a}_{\mu j+1\sigma}+\hat{a}^\dagger_{\mu j\sigma}\hat{a}_{\mu j-1\sigma}\right)
%+
%\frac{J_c}{4}\sum\limits_{i\ne j\ne k=1}^{3}\sum_{\sigma,\sigma'} \hat{a}_{\mu i\sigma} \hat{a}^\dagger_{\mu j\sigma} (1-\hat n^{(a)}_{\mu j\uparrow})(1-\hat n^{(a)}_{\mu j\downarrow})\hat a_{\mu j\sigma'}\hat a^\dagger_{\mu k\sigma'}
% \right]P_0
% \label{HtJ}
%\end{eqnarray}
%\end{widetext}
where $\hat{a}^\dagger_{\mu j\sigma}$ creates an electron with spin $\sigma$ in the $j$th Wannier orbital, $\hat n^{(x)}_{\{y\}}=\hat x^\dagger_{\{y\}}\hat x_{\{y\}}$, and $P_0$ projects out states that contain empty sites.
In an effective low energy theory of the Hubbard model $J_c=4t_c^2/U +{\mathcal O}(t_c^3/U^2)$.  
Note that the second term in Eq. (\ref{HtJ}) retains the `three site' terms that are often neglected near half-filling \cite{Essler}, as that limit will not be uniformly applicable below.

The  SMOC  is most naturally written in terms of the `Condon-Shortley' operators \cite{Amie16}, $\hat c_{\mu k\sigma}$, that create an electron on molecule $\mu$ with angular momentum $L_z=k$ about the $z$-axis and spin $\sigma$:
%\begin{subequations}
	\begin{eqnarray}
	\hat c_{\mu k\sigma}&=&\frac{i^ki^{|k|}}{\sqrt{3}}\sum_{j=1}^3 \hat a_{\mu j\sigma}^\dagger e^{i\phi_jk}, \label{CS}
	%		\hat c_{\mu 1\sigma}&=&-\frac{1}{\sqrt{3}}\sum_{j=1}^3 \hat a_{\mu j\sigma}^\dagger e^{i\phi_j}\\
%		\hat c_{\mu 0\sigma}&=&\frac{1}{\sqrt{3}}\sum_{j=1}^3 \hat a_{\mu j\sigma}^\dagger \\
%		\hat c_{\mu -1\sigma}&=&\frac{1}{\sqrt{3}}\sum_{j=1}^3 \hat a_{\mu j\sigma}^\dagger e^{-i\phi_j},
	\end{eqnarray}
%	\label{CS}
%\end{subequations}
where $\phi_j=2\pi(j-1)/3$.
Thus one finds \cite{Amie16,Jaime16} that
%\begin{widetext}
\begin{subequations}
\begin{eqnarray}
H_{SMO}^{(\mu)}&\equiv&\lambda^{(\mu)}_z\hat L^z_\mu\hat S^z_\mu +\frac{\lambda^{(\mu)}_{xy}}{2}\left(\hat L^-_\mu\hat S^+_\mu+\hat L^+_\mu\hat S^-_\mu\right)\label{HSOa}\\
&=&\frac{\lambda^{(\mu)}_z}{2} \left(\hat n^{(c)}_{\mu 1\uparrow} - \hat n^{(c)}_{\mu 1\downarrow} -\hat n^{(c)}_{\mu -1\uparrow} + \hat n^{(c)}_{\mu -1\downarrow} \right)
% \\ && \notag
+
\frac{\lambda^{(\mu)}_{xy}}{\sqrt{2}}\left(
\hat c^\dagger_{\mu -1\uparrow}\hat c_{\mu 0\downarrow}
+\hat c^\dagger_{\mu 0\uparrow}\hat c_{\mu 1\downarrow} 
+\hat c^\dagger_{\mu 1\downarrow}\hat c_{\mu 0\uparrow}
+\hat c^\dagger_{\mu 0\downarrow}\hat c_{\mu -1\uparrow}\right)
\\
&=&\frac{i\lambda^{(\mu)}_z}{3}\sum_{j\ell} \left(\hat a^\dagger_{\mu j\uparrow}a_{\mu \ell\uparrow} 
- \hat a^\dagger_{\mu j\downarrow}a_{\mu \ell\downarrow}  \right)\sin\phi_{j-\ell}
%\notag\\&&
+
\frac{\lambda^{(\mu)}_{xy}}{3\sqrt{2}}
\left[
\hat a^\dagger_{\mu j\uparrow}a_{\mu \ell\downarrow}\left( e^{-i\phi_j}-e^{-i\phi_\ell} \right) 
%\right.
%\notag \\ && \left. \hspace{1cm}
- \hat a^\dagger_{\mu \ell\downarrow}a_{\mu j\uparrow}\left( e^{i\phi_\ell}-e^{i\phi_j} \right) \right] \label{HSOreal}
\end{eqnarray}
\label{HSO}
\end{subequations}

In this paper, we will primarily be interested in the DM interaction, which has leading terms at linear order in the SOC. Therefore, we analyze the Hamiltonian $H_{tJ}+H_{SMO}$, where $H_{tJ}=\sum_{\mu}H_{tJ}^{(\mu)}$ and $H_{SMO}=\sum_{\mu}H_{SMO}^{(\mu)}$, via first order perturbation theory below, with $H_{SMO}$ taken as the perturbation.

\Mo is two-thirds filled (i.e., an average of two holes per molecule).
Furthermore, we will see below that there is a strong analogy between the molecular problem and the Hund's physics at play in many transition metal oxides \cite{Perkins,GeorgesARCMP}. For a $t_{2g}$ orbital near the Fermi energy the most interesting effects occur at the generic fillings, two electrons or two holes in the three orbitals {--} half-filling (three electrons/holes) and one electron or hole per molecule are special cases.  Furthermore, in the molecular case with $t_c>0$ the two-electron case is a trivial band insulator. Therefore, we focus on the two-thirds filled insulator below.

\begin{figure*}
	\begin{center}
		\includegraphics[width=0.32\columnwidth]{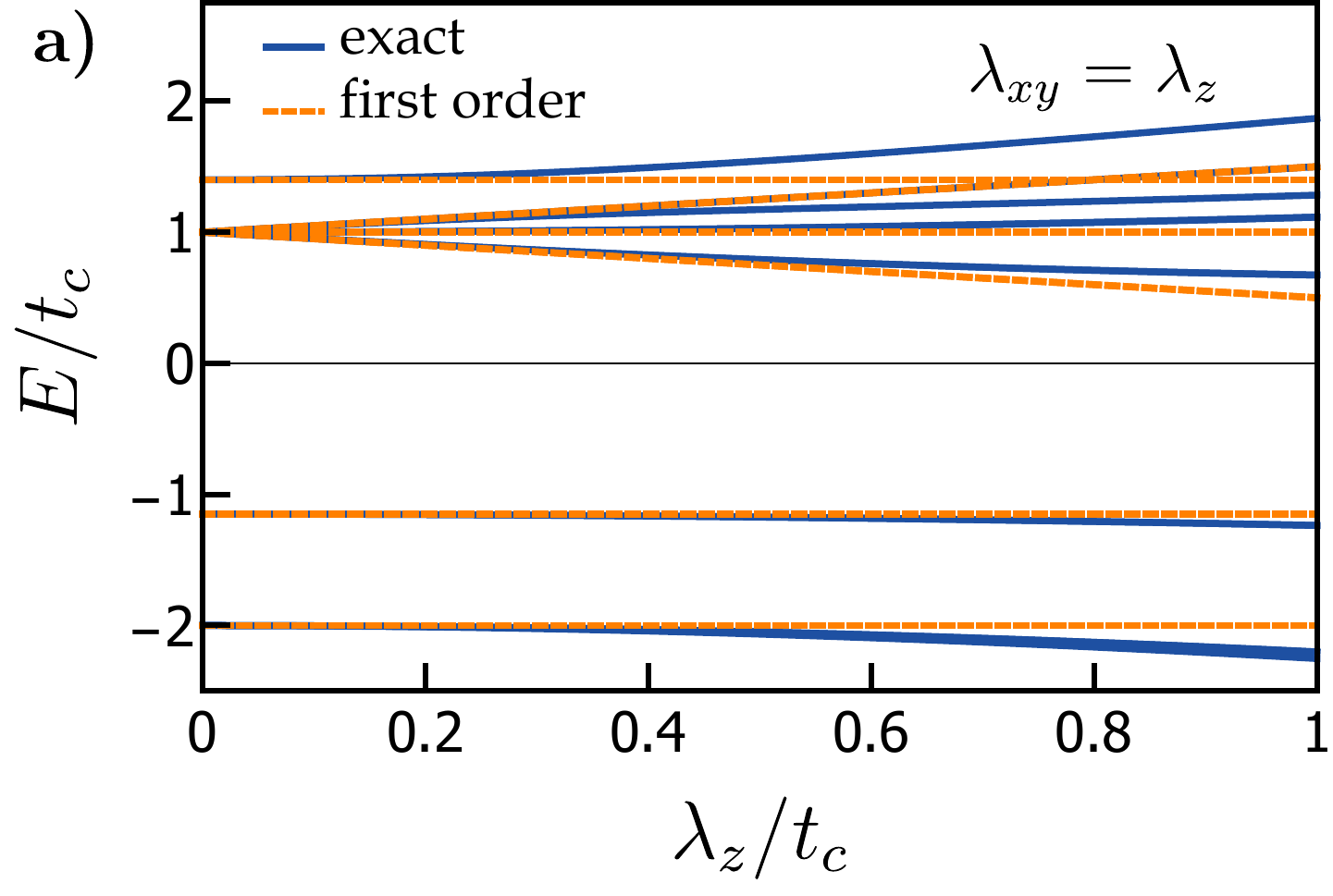}
		\includegraphics[width=0.32\columnwidth]{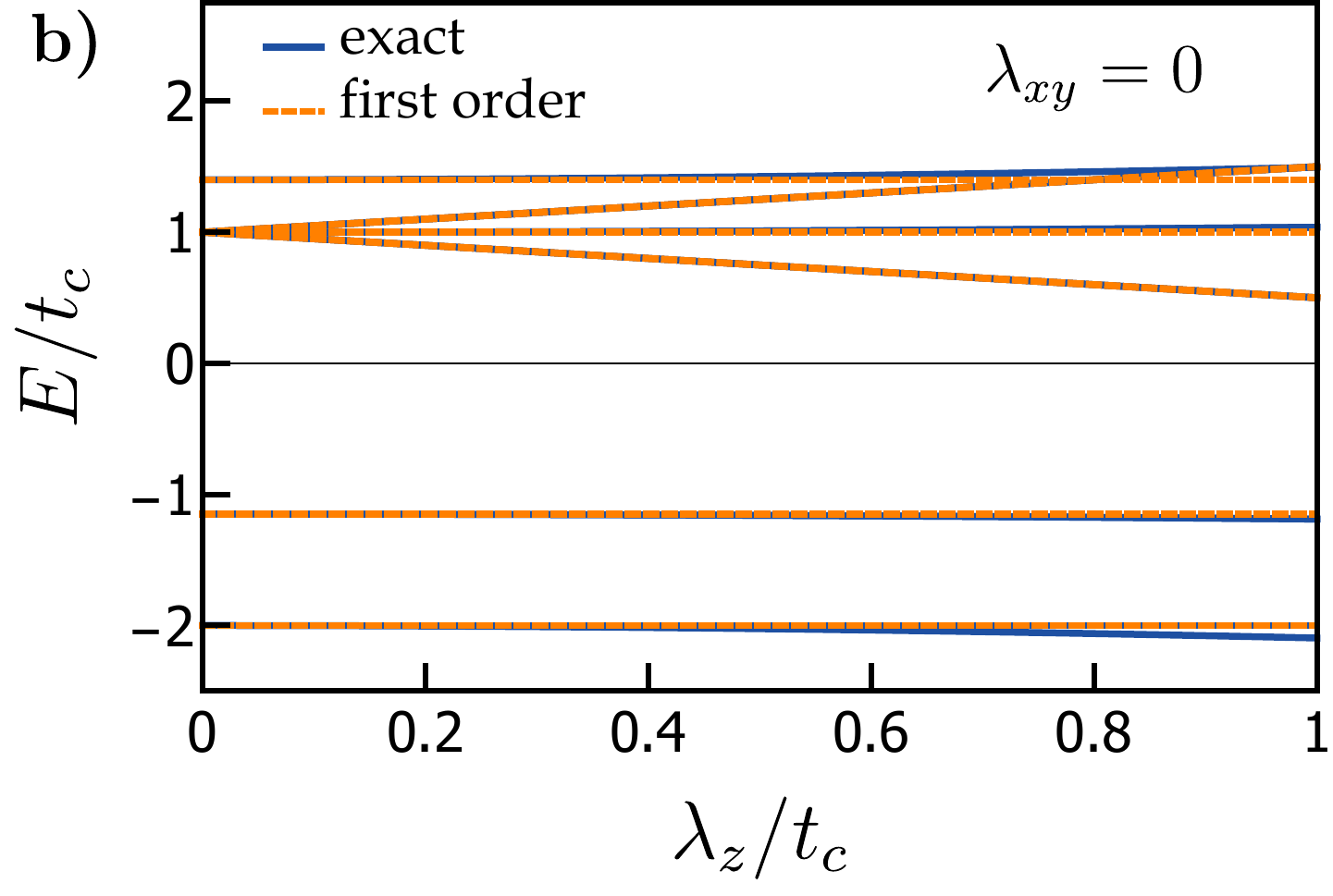}
		\includegraphics[width=0.32\columnwidth]{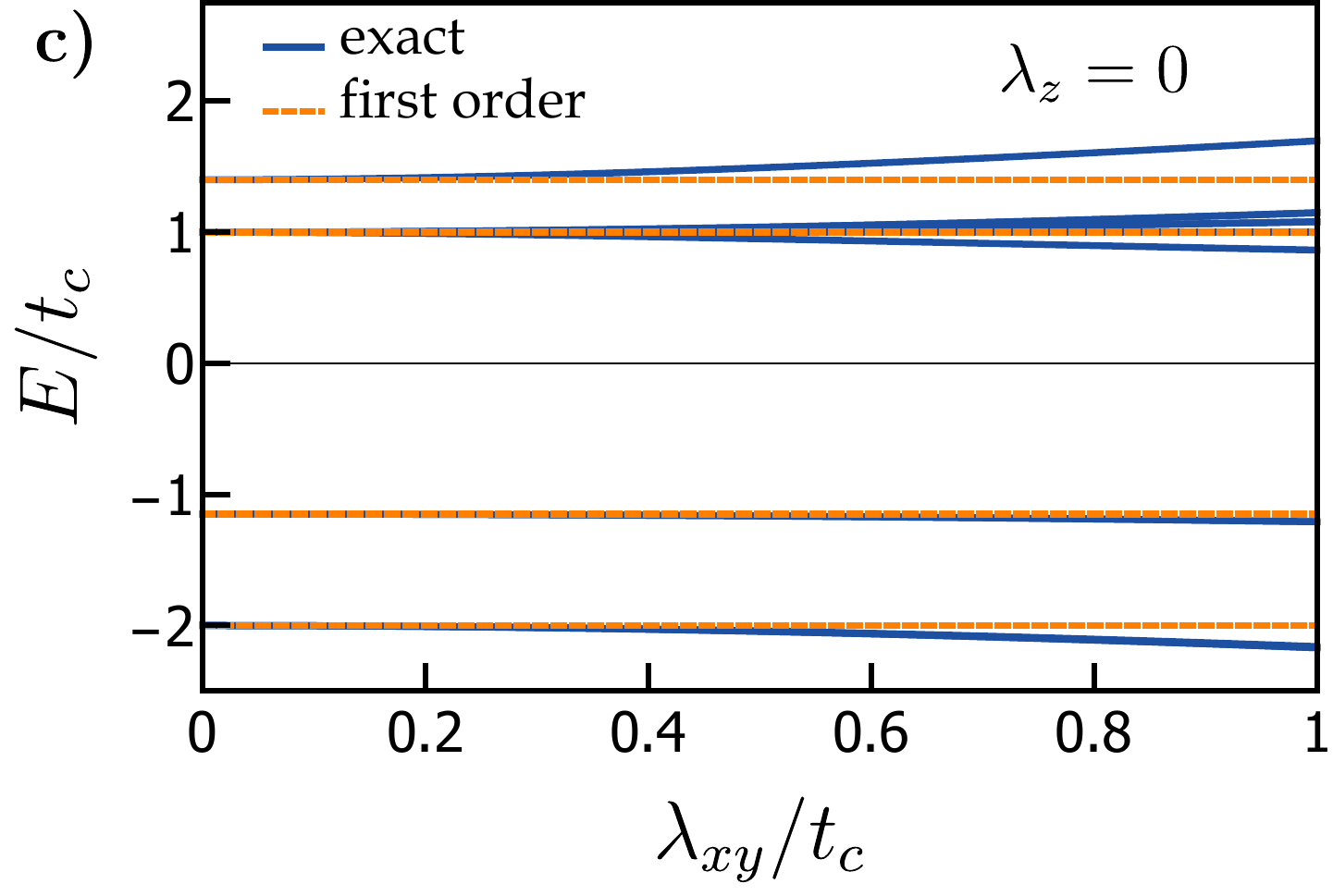}
	\end{center}
	\caption{Comparison of the exact solution for a single molecule with the energy to first order in the SMOC. For small $J_c$ the spectrum is not strongly affected by the value of $J_c$; here we set $J_c=0.3t_c$ in all panels. The perturbation theory is most accurate for $\lambda_{xy}=0$. This is unsurprising because, at first order, there are  corrections to the energies due to $\lambda_{z}$ but not $\lambda_{xy}$. This is natural because in the absence of SMOC Bloch's theorem requires that the single molecule energy eigenstates are also eigenstates of ${\hat L}_\mu^z$. 
	}
	\label{fig:spectra}
\end{figure*}

%\subsection{Neutral molecule}   

At first order in the SMOC the four electron ground state is three-fold degenerate in the physically relevant parameter regime $0<2J_c<t_c$, with energy $E_4^{3A}=-2 t_c$. Henceforth we  label the energies  $E_n^{(2S+1)\Gamma}$, where $n$ is the number of electrons, $S$ is the magnitude of spin in the absence of SMOC, and $\Gamma\in\{A,E\}$ labels the representation of the orbital part in the absence of SMOC; $A$ ($E$) irreducible representations require singly (doubly) orbitally degenerate  wavefunctions.
Similarly, we will label the eigenstates $\ket{\Phi_n^{(2S+1)\Gamma}(j)}$, where $j$ runs over the degenerate states, and is somewhat analogous to {a projection of} the total angular momentum in the spherically symmetric case. 
 The ground state wavefunctions, $\ket{\Phi_4^{3A}(j)}$, are 
\begin{subequations}
	\begin{eqnarray}
\ket{\Phi_4^{3A}(1)}	=%&=& 
	\frac{1}{\sqrt{3}}\sum_{j} \bigg[ 
	\hat{a}_{\mu j\downarrow} \hat{a}_{\mu j-1\downarrow}
%\notag \hspace*{3.6cm}\\
	-\frac{i\lambda_{xy}e^{-i\phi_{j+1}}}{\sqrt{6} (J_c - 2 t_c)} \left( \hat{a}_{\mu j\uparrow} \hat{a}_{\mu j-1\downarrow} - \hat{a}_{\mu j\downarrow} \hat{a}_{\mu j-1\uparrow}  \right)
\notag \hspace*{0.7cm}\\%&& \hspace{1cm} 
	-\frac{\lambda_{xy}}{6\sqrt{2} t_c} e^{-i\phi_{j+1}} \left( \hat{a}_{\mu j\uparrow} \hat{a}_{\mu j-1\downarrow} + \hat{a}_{\mu j\downarrow} \hat{a}_{\mu j-1\uparrow}  \right)
	\bigg] \ket{vac_h}, 
	%\notag\\
	%\\
\end{eqnarray}
\begin{eqnarray}
\ket{\Phi_4^{3A}(0)}	=%&=& 
	-\frac{1}{\sqrt{6}}\sum_{j} \bigg[ 
	\left( \hat{a}_{\mu j\uparrow} \hat{a}_{\mu j-1\downarrow} + \hat{a}_{\mu j\downarrow} \hat{a}_{\mu j-1\uparrow}  \right)
%	\hspace*{1.2cm}
%\notag \\
	-\frac{\lambda_z}{\sqrt{3} (2 J_c - 4 t_c)}  \left( \hat{a}_{\mu j\uparrow} \hat{a}_{\mu j-1\downarrow} - \hat{a}_{\mu j\downarrow} \hat{a}_{\mu j-1\uparrow}  \right) 
\notag \\
	+\frac{\lambda_{xy}}{6 t_c}  e^{i\phi_{j+1}} \hat{a}_{\mu j\downarrow} \hat{a}_{\mu j-1\downarrow} 
%	\hspace*{3.35cm}
%	\notag\\
	+\frac{\lambda_{xy}}{6 t_c}  e^{-i\phi_{j+1}} \hat{a}_{\mu j\uparrow} \hat{a}_{\mu j-1\uparrow}  
 \bigg]\ket{vac_h}, \hspace*{1.85cm}
\end{eqnarray}
and
\begin{eqnarray}
	\ket{\Phi_4^{3A}(-1)}	=%&=& 
	\frac{1}{\sqrt{3}}\sum_{j} \bigg[ 
	\hat{a}_{\mu j\uparrow} \hat{a}_{\mu j-1\uparrow}
%\notag \hspace*{3.6cm}\\
	+\frac{i\lambda_{xy}e^{i\phi_{j+1}}}{\sqrt{6} (J_c - 2 t_c)}  \left( \hat{a}_{\mu j\uparrow} \hat{a}_{\mu j-1\downarrow} - \hat{a}_{\mu j\downarrow} \hat{a}_{\mu j-1\uparrow}  \right) 
	\hspace*{0.55cm}
\notag\\%&& \hspace{1cm}
	 \left.
	 + \frac{\lambda_{xy}}{6\sqrt{2} t_c}  e^{i\phi_{j+1}} \left( \hat{a}_{\mu j\uparrow} \hat{a}_{\mu j-1\downarrow} + \hat{a}_{\mu j\downarrow} \hat{a}_{\mu j-1\uparrow}  \right) 
	\right] \ket{vac_h},
	\notag\\
	\end{eqnarray}
\end{subequations}
\end{widetext}
where
$\ket{\textrm{vac}_h}\equiv\hat a^\dag_{\mu 1\downarrow}\hat a^\dag_{\mu 2\downarrow}\hat a^\dag_{\mu 3\downarrow}\hat a^\dag_{\mu 1\uparrow}\hat a^\dag_{\mu 2\uparrow}\hat a^\dag_{\mu 3\uparrow}\ket{0}$ is the vacuum for holes and we have suppressed the molecular indices, $\mu$, on $\lambda^{(\mu)}_{xy}$ and $\lambda^{(\mu)}_{z}$ for clarity (as we will do henceforth when the context is clear).

The excited states are described in Appendix \ref{sect:neut}. In Fig. \ref{fig:spectra} we compare the first order spectrum with the exact solution of the three-site \tJ model with SMOC. We see that for weak to moderate SMOC the first order expressions provide an adequate description of the single molecule spectrum. Furthermore, we see that there is a large gap ($t_c-J_c/2$) to the first excited states, $\ket{\Phi_4^{1E}(\pm1)}$. {Note that the \tJ model is only valid in the limit $U\gg t_c$, which implies that $t_c\gg J_c\simeq 4t_c^2/U$. So this gap cannot close in the regime in which our current treatment in valid. Numerical investigations of the Hubbard model with SMOC on three sites \cite{Jaime16,Jaime17} find that this gap does not close for reasonable values of $U$, unless an explicit antiferromagnetic exchange interaction between neighboring Wanniers on the same molecule is also included. }

In the absence of SMOC the ground states reduce to a spin-triplet ($S^z\in\{-1,0,1\}$) with angular momentum $k=0$ about the $z$-axis. SMOC mixes these states with higher lying states with the same $j=k+S^z$. Here there is an important difference from the atomic case: the addition of angular momentum occurs modulo three onto the interval $(-3/2,3/2]$.
At linear order in the SMOC the ground state is also a spin triplet. This changes at second order: nevertheless the low-energy physics can still be understood in terms of a pseudospin, $\bm{\mathcal S}$, triplet. At second order the degeneracy is lifted by a trigonal splitting   of the triplet: $\Delta({\mathcal S}^z)^2$, where $\Delta=[{2t_c\lambda_z^2}-{J_c \lambda_{xy} ^2}]/{12 (2 t_c-J_c)}$; generically one expects $\Delta>0$ as $t_c\gg J_c$, however it is possible to have $\Delta<0$ due to the anisotropy in the SMOC if $(\lambda_{xy}/\lambda_z)^2>t_c/J_c\simeq U/4t_c$. Furthermore the Heisenberg exchange interactions become spatially anisotropic. We will not discuss second order effects further here -- see  \cite{Jaime16,Jaime17} for details. 

Note that in specifying the eigenstates above one has picked an explicit relative gauge in the effective low-energy Hamiltonian. The above choice simplifies the analysis as it ensures that the spin-one Pauli matrices take their usual form. 

The  physics of the singly charged cation and anion, which will be required to derive the effective low-energy spin models, below, are described in Appendices \ref{append3e} and \ref{append5e} respectively.

\section{Effective interaction between molecules}\label{sect:2mol-SMOC}

We now consider a variety of potential couplings between molecules within the context of the \tJ model. Intramolecular couplings are considered within perturbation theory. Thus, we derive  effective low-energy Hamiltonians for the interactions between the pseudospin-one moments in the ground state of the trimer with two holes.  
We use DiracQ  \cite{DiracQ} package for Mathematica to evaluate the matrix elements for the effective spin-one model, using the wavefunctions calculated to first order in SMOC (Section \ref{sect:1mol} and Appendices \ref{sect:neut}-\ref{append5e}). 

{In this section we consider both packing motifs relevant to \Mo and  natural generalizations of these packing motifs, which may be realized in other trinuclear complexes. }
	
{
	Firstly, we discuss molecule coupled by hopping between a single orbital on each molecule. The internal structure of the molecule means that the \DM coupling between pairs of molecules related by inversion (Fig. \ref{fig:lattice}c; section \ref{sect:inv}) is significantly different from that for pairs related by $\pi$-rotations (Fig. \ref{fig:lattice}d; section \ref{sect:rot}). The former case is relevant to \Mo and the later emphasizes the additional physics due to the internal structure of molecular crystals. It is also natural to consider what happens when these symmetries are broken, which we do in sections \ref{sect:brokenI} and  \ref{sect:bC2}.
}

{
	Secondly, we discuss molecules coupled by hopping from each Wannier orbital to the equivalent Wannier orbital on a neighboring molecule (cf. Fig. \ref{fig:lattice}b) in \ref{sect:interlayer}. This is the case relevant to interlayer hopping in \Mo.
}

\subsection{In-plane coupling}\label{sect:inplane}

We first consider a pair of molecules coupled through a single hopping integral, $t_g$, as sketched in Fig. \ref{fig:lattice}c. For example, this is the strongest in-plane coupling between molecules in \Mo \cite{JackoPRB15}{, see Table \ref{tab:Jacko}}.

We first consider an intermolecular hopping, which, without loss of generality, we take to couple the Wannier orbital labeled ``1'' on each molecule:
\begin{eqnarray}
H_{t_g}=-t_g\sum_{\langle\mu\nu\rangle\sigma} P_0\left( 
\hat a_{\mu1\sigma}^\dag \hat a_{\nu1\sigma} +
\hat a_{\mu1\sigma}^\dag \hat a_{\nu1\sigma} \right)P_0. \label{HamG}
\end{eqnarray}

For $t_g=0$ the ground state is nine-fold degenerate (as the ground state of a single molecule is three-fold degenerate).
Making a canonical transformation one finds that, to second order in $t_g$, the effective interaction between the degenerate ground states of the $t_g=0$ problem is  described by
\begin{eqnarray}
&&\hspace*{-1.5cm}\bra{\tilde{\mathcal S}_\mu^z,\tilde{\mathcal S}_\nu^z}H^\textrm{eff}\ket{{\mathcal S}_\mu^z,{\mathcal S}_\nu^z} \notag\\
&\equiv&
-\sum_n\frac{ \bra{\tilde{\mathcal S}_\mu^z}\bra{\tilde{\mathcal S}_\nu^z}H_{t_g}|n\rangle\langle n|H_{t_g} \ket{{\mathcal S}_\mu^z}\ket{{\mathcal S}_\nu^z}}{E_n-2E_4^{3A}}\notag\\
&=&
-t_g^2\sum_{n_3=1}^8\sum_{n_5=1}^6 \sum_{\sigma,\sigma'} \frac{1}{E_{n_3}+E_{n_5}-2E_4^{3A}}\notag\\
&&
\hspace*{0cm}\times\Big( 
 \bra{\tilde{\mathcal S}_\mu^z}
\hat a_{\mu 1\sigma}^\dag 
\ket{n_3^\mu}
\bra{n_3^\mu}
\hat a_{\mu 1\sigma'} 
\ket{{\mathcal S}_\mu^z}
\notag\\
&& \hspace*{.75cm}\times
\bra{\tilde{\mathcal S}_\nu^z}
\hat a_{\nu 1\sigma} 
\ket{n_5^\nu}
\bra{n_5^\nu}
\hat a_{\nu 1\sigma'}^\dag 
\ket{{\mathcal S}_\nu^z} 
\notag\\
&& \hspace*{.5cm} +
\bra{\tilde{\mathcal S}_\mu^z}
\hat a_{\mu 1\sigma}
\ket{n_5^\mu}
\bra{n_5^\mu}
\hat a_{\mu 1\sigma'}^\dag  
\ket{{\mathcal S}_\mu^z}
\notag\\
&& \hspace*{.75cm}\times
\bra{\tilde{\mathcal S}_\nu^z}
\hat a_{\nu 1\sigma}^\dag 
\ket{n_3^\nu}
\bra{n_3^\nu}
\hat a_{\nu 1\sigma'} 
\ket{{\mathcal S}_\nu^z}
\Big)%\\
\label{eqn:MEt}
\end{eqnarray}
where ${\mathcal S}^z_\mu$ ($\tilde{\mathcal S}^z_\mu$) $\in\{-1,0,1\}$ is the initial (final) spin on the $\mu$th molecule, i.e., $\ket{{\mathcal S}^z_\mu}\equiv\ket{\Phi_4^{3A}({\mathcal S}^z_\mu)}$, and $n_3$ ($n_5$) runs over all eigenstates of the three (five) electron monomer problem, see Appendix \ref{append3e} (\ref{append5e}). 

As the \tJ model is derived from the Hubbard model at lowest order in $t/U$ only including the terms described by Eq. (\ref{eqn:MEt}) neglects virtual transitions that change the occupation of the ``1'' orbitals on either molecule. To correct this we also include a  Heisenberg coupling between the 1-Wannier orbitals on the monomers. As there are no `three site' terms relevant to this interaction the relevant term in the Hamiltonian is
\begin{eqnarray}
H_{J_g}=\sum_{\langle\mu\nu\rangle} J_gP_0\left( {\bm S}_{\mu1}\cdot{\bm S}_{\nu1} -\frac{\hat n_{\mu1}^{(h)}\hat n_{\nu1}^{(h)}}{4} \right)
 P_0, \label{HamJG}
\end{eqnarray}
where, $\hat n_{\mu i}^{(x)}=\sum_{\sigma} \hat n_{\mu i\sigma}^{(x)}$, ${\bm S}_{\mu i}=\sum_{\alpha\beta} \hat h_{\mu i\alpha}^\dagger {\bm \sigma}_{\alpha\beta} \hat h_{\mu i\beta}$ is the spin operator for holes, $\hat h_{\mu i\alpha}^\dagger =\hat a_{\mu i\alpha}$, and, in the Hubbard model, $J_g=4t_g^2/U + {\mathcal O}(t^4/U^3)$ and hence $J_g/J_c=(t_g/t_c)^2+\dots$. As $J_g$ is already an effect at second order in $t_g$, we only consider perturbations at first order in $J_g$ for consistency. 
%The relevant matrix elements are
%\begin{eqnarray}
%\bra{S_f^A}\bra{S_f^B}H_{J_g}\ket{S_i^B}\ket{S_i^A}
%&=& 
%%J_c\bra{S_f^A}\bra{S_f^B}\left( S^z_{A1} S^z_{B1} + \frac{S^+_{A1} S^-_{B1}}{2} + \frac{S^-_{A1} S^+_{B1}}{2} -\frac{\hat n_{A1}\hat n_{B1}}{4} \right)\ket{S_i^B}\ket{S_i^A}
%%\\&=&
%J_c\left( 
%\bra{S_f^A} S^z_{A1}\ket{S_i^A} \bra{S_f^B}S^z_{B1}\ket{S_i^B} 
%+ \frac{\bra{S_f^A}S^+_{A1}\ket{S_i^A} \bra{S_f^B}S^-_{B1}\ket{S_i^B}}{2}
%\right.\notag\\&&\left. \hspace*{1cm}
%+ \frac{\bra{S_f^A}S^-_{A1}\ket{S_i^A} \bra{S_f^B}S^+_{B1}\ket{S_i^B}}{2} 
%-\frac{\bra{S_f^A}\hat n_{A1}\ket{S_i^A} \bra{S_f^B} \hat n_{B1}\ket{S_i^B}}{4} 
%\right). \label{eqn:MEJ}
%\end{eqnarray}

\subsubsection{Molecules related by inversion symmetry}\label{sect:inv}

{This is the case relevant to nearest neighbors in the plane in \Mo.}

As angular momenta [including spin and the molecular angular momentum, $\bm L_\mu$ cf. Eq. (\ref{HSOa})] are pseudovectors $H_{SMO}^{(\mu)}$ must be identical for two molecules related by inversion. Therefore we set $\lambda_{xy}^{\mu}=\lambda_{xy}$ and $\lambda_z^{\mu}=\lambda_z$ for all molecules.

We find that, to linear order in $H_{SMO}$, SMOC has no effect and the effective Hamiltonian is
\begin{eqnarray}
H_{{\mathcal S}=1}^\text{eff}=-2Nt_c+\sum_{\mu\nu}{\mathcal J}_{\mu\nu}\left( \bm{\mathcal S}_\mu\cdot\bm{\mathcal S}_\nu -  \frac{\hat n_\mu^{(h)} \hat n_\nu^{(h)}}{4} \right)
\label{HnoDM}
\end{eqnarray}
where $N$ is the number of molecules, $\hat n_{\mu}^{(x)}=\sum_{i} \hat n_{\mu i}^{(x)}$ and if $\mu$ and $\nu$ are nearest neighbors in the plane (i.e., with the intermolecular coupling through a single orbital per molecule, as described by Eqs. (\ref{HamG}) and (\ref{HamJG}), cf. Fig. \ref{fig:lattice}c) the effective Heisenberg exchange constant is ${\mathcal J}_{\mu\nu}={\mathcal J}_\|$, where
\begin{eqnarray}
\label{EqJ}
{\mathcal J}_\|=\frac{J_g}{9}+\frac{4 J_c t_g^2}{81 (2 t_c-J_c ) t_c}. 
\end{eqnarray}

It is not surprising that there is no DM interaction for the inversion symmetric case \cite{Moriya}:  ${\mathcal I}{\bm D}\cdot\bm{\mathcal S}_\mu\times\bm{\mathcal S}_\nu{\mathcal I}^{-1}={\bm D}\cdot\bm{\mathcal S}_\nu\times\bm{\mathcal S}_\mu=-{\bm D}\cdot\bm{\mathcal S}_\mu\times\bm{\mathcal S}_\nu$, where $\mathcal I$ is the inversion operator. Thus, inversion requires that ${\bm D}={\bm 0}$.

\subsubsection{Broken inversion symmetry} \label{sect:brokenI}

If there is no symmetry relation between molecules, e.g., they sit in crystallographically distinct locations, then there is no {\it a priori} relationship between the intra-molecular terms in the Hamiltonian. Nevertheless for otherwise identical molecules the differences may  be small. Therefore, in this section we assume that the intra-molecular Hamiltonians are the same for all molecules and explore the direct consequences of molecular packing on the effective pseudospin model.

A natural approach to this is to start from the inversion symmetric problem and rotate one of the molecules around some axis. 
%This presents an immediate problem: it is obvious from the definitions above that the $z$-axis is perpendicular to the plane defined by the three Wannier orbitals, but where are the $x$- and $y$-axes?
%
One is not free to choose the $x$- and $y$-axes arbitrarily as, in specifying the form of the SMOC, we  have implicitly defined the coordinate system. %Therefore, the problem is reduced to extracting this information from the SOC term in the Hamiltonian. 
In particular, it follows from the definitions given in Eqs. (\ref{CS}) and (\ref{HSO}) that 
\begin{eqnarray}
L^+_\mu&=&\frac{\sqrt2}3\sum_{lj\sigma}\hat a_{\mu l\sigma}^\dag\hat a_{\mu j\sigma}\left(e^{i\phi_j}-e^{i\phi_l}\right),\\
L^-_\mu&=&\frac{\sqrt2}3\sum_{lj\sigma}\hat a_{\mu l\sigma}^\dag\hat a_{\mu j\sigma}\left(e^{-i\phi_l}-e^{-i\phi_j}\right),\\
L^x_\mu&=&\frac1{2}\left(L^+_\mu+L^-_\mu\right)
\notag\\
&=&\frac{i{\sqrt2}}3 \sum_{lj\sigma}\hat a_{\mu l\sigma}^\dag\hat a_{\mu j\sigma}\left(\sin\phi_j-\sin\phi_l\right),
\\
L^y_\mu&=&\frac{-i}{2}\left(L^+_\mu-L^-_\mu\right)
\notag\\
&=&\frac{i{\sqrt2}}3 \sum_{lj\sigma}\hat a_{\mu l\sigma}^\dag\hat a_{\mu j\sigma}\left(\cos\phi_j-\cos\phi_l\right).
\end{eqnarray}
Thus the $x$-axis is parallel to $\bm R_{\mu3}-\bm R_{\mu2}$ (i.e., perpendicular to the intermolecular bond link{ing} the sites labeled 1) and the $y$-axis is parallel to $2\bm R_{\mu1}-\bm R_{\mu3}-\bm R_{\mu2}$ (i.e., parallel to the intermolecular bond {linking the sites labeled 1}), cf. Fig. \ref{fig:lattice}a.  

As we are dealing with spin-1/2 particles we rotate by an angle $\vartheta$   about the $\bm{\hat\vartheta}$  axis on the $\mu$th molecule by applying the unitary transformation
\begin{eqnarray}
{\mathcal R}^{(1/2)}_{\mu}({\bm\vartheta})=\exp\left( -\frac{i\sum_{i\alpha\beta}\hat a_{\mu i\alpha}^\dag\bm{\hat\vartheta}\cdot{\bm\sigma}_{\alpha\beta}\hat a_{\mu i\beta}}{2} \right) \label{eqn:trans}
\end{eqnarray}
where ${\bm\vartheta}=\vartheta\bm{\hat\vartheta}$ and ${\bm\sigma}$ is the vector of Pauli matrices. 
%Explicitly we have
%\begin{subequations}\label{eqn:trans}
%\begin{eqnarray}
%\left[ {\mathcal R}^{x}_\mu(\theta_x) \right]^\dag\hat a_{\mu i\sigma}{\mathcal R}^{x}_\mu(\theta_x)&=&\hat a_{\mu i\sigma}\cos\frac{\theta_x}2 - i\hat a_{\mu i\overline\sigma}\sin\frac{\theta_x}2 \\
%\left[ {\mathcal R}^{x}_\mu(\theta_x) \right]^\dag\hat a_{\mu i\sigma}^\dag{\mathcal R}^{x}_\mu(\theta_x)&=&\hat a_{\mu i\sigma}^\dag\cos\frac{\theta_x}2 + i\hat a_{\mu i\overline\sigma}^\dag\sin\frac{\theta_x}2 \\
%\left[ {\mathcal R}^{y}_\mu(\theta_y) \right]^{\dag}\hat a_{\mu i\sigma}^{(\dag)}{\mathcal R}^{y}_\mu(\theta_y)&=&\hat a_{\mu i\sigma}^{(\dag)}\cos\frac{\theta_y}2 - 2\sigma\hat a_{\mu i\overline\sigma}^{(\dag)}\sin\frac{\theta_y}2 \\
%\left[ {\mathcal R}^{z}_\mu(\theta_z) \right]^{\dag}\hat a_{\mu i\sigma}^{\dag}{\mathcal R}^{z}_\mu(\theta_z)&=&\hat a_{\mu i\sigma}^{\dag}e^{i\theta_z\sigma}  \\
%\left[ {\mathcal R}^{z}_\mu(\theta_z) \right]^{\dag}\hat a_{\mu i\sigma}{\mathcal R}^{z}_\mu(\theta_z)&=&\hat a_{\mu i\sigma}e^{-i\theta_z\sigma}.
%\end{eqnarray} 
%\end{subequations}
%Note that for consistency with the above calculations we take $\sigma=\pm1/2$ in the above equations.

\begin{figure}
	\begin{center}
		\includegraphics[width=0.9\columnwidth]{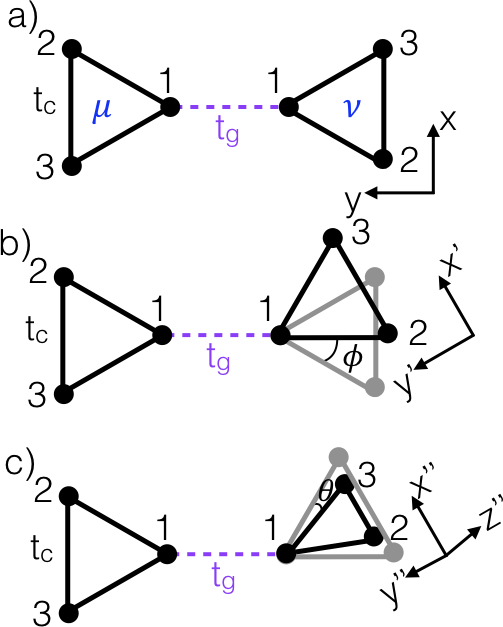}
	\end{center}
	\caption{
		{
			Molecules packing with a rotation relative to one another leads to a DM coupling. (a) Initial arrangement with inversion symmetry and hence no DM interaction. The local axes of the $\nu$th molecule are marked (the $z$-axis is perpendicular to the page.) (b) The $\nu$th molecule is rotated by and angle $\phi$ about the $z$-axis, defining a new local coordinate system, $(x',y',z')$. (c) The $\nu$th molecule is rotated by an angle $\theta$ about the $x'$-axis, defining a new local coordinate system, $(x'',y'',z'')$. To produce an arbitrary,  rotation, ${\bm\vartheta}$, one must also complete another rotation about the $z''$,  not shown here. Translations have a less complicated effect on the effective Hamiltonian (only  via changes the parameters of the microscopic Hamiltonian) are therefore are not shown in this figure.
		}
	}
	\label{fig:latticeRot}
\end{figure}

The simplest method to implement this is to work always in the local frame of each molecule, this requires that we perform the inverse transformation on the intermolecular terms in the Hamiltonian. For concreteness we take the rotation to occur on the $B$ molecule. %Hence,
%\begin{eqnarray}
%H_{t_g}&\rightarrow& \left[ {\mathcal R}^{(1/2)}_{B}(-{\bm\vartheta})\right]^\dag H_{t_g} {\mathcal R}^{(1/2)}_{B}(-{\bm\vartheta})
%\notag\\
%&=&-t_g\sum_{\sigma} P_0\left( 
%\hat a_{A1\sigma}^\dag \left[ {\mathcal R}^{(1/2)}_{B}(-{\bm\vartheta})\right]^\dag \hat a_{B1\sigma} {\mathcal R}^{(1/2)}_{B}(-{\bm\vartheta}) \right.
%\notag\\
%&& + \left.
%\left[ {\mathcal R}^{(1/2)}_{B}(-{\bm\vartheta})\right]^\dag \hat a_{B1\sigma}^\dag {\mathcal R}^{(1/2)}_{B}(-{\bm\vartheta}) \hat a_{A1\sigma} \right)P_0
%\end{eqnarray}
We represent an arbitrary rotation by the Euler angles, $\phi$, $\theta$, and $\psi$ taken to be about the $z$, $x'$, and $z''$ axes respectively, where the primed (double primed) coordinate system is that of the molecule after one (two) rotations{, as sketched in Fig. \ref{fig:latticeRot}}. Setting ${\mathcal R}^{(1/2)}_{B}(-{\bm\vartheta}) = {\mathcal R}^{(1/2)}_{B}(\phi\bm{z}) {\mathcal R}^{(1/2)}_{B}(\theta\bm{x'}) {\mathcal R}^{(1/2)}_{B}(\psi\bm{z''})$ yields
\begin{eqnarray}
&&\left[ {\mathcal R}^{(1/2)}_{B}(-{\bm\vartheta})\right]^\dag  H_{t_g} {\mathcal R}^{(1/2)}_{B}(-{\bm\vartheta})
%-t_g\sum_{\sigma} P_0\left( 
%\hat a_{A1\sigma}^\dag 
%\left[ {\mathcal R}^{x}_{B}(\theta) \right]^\dag
%\left[ {\mathcal R}^{z}_{B}(\phi) \right]^\dag 
%\hat a_{B1\sigma} 
%{\mathcal R}^{z}_{B}(\phi) {\mathcal R}^{x}_{B}(\theta) \right. \notag\\
%&&\hspace{2cm}+  \left.
%\left[ {\mathcal R}^{x}_{B}(\theta) \right]^\dag
%\left[ {\mathcal R}^{z}_{B}(\phi) \right]^\dag 
%\hat a_{B1\sigma}^\dag 
%{\mathcal R}^{z}_{B}(\phi) {\mathcal R}^{x}_{B}(\theta) 
%\hat a_{A1\sigma} \right)P_0\\&=&
%-t_g\sum_{\sigma} P_0\left( 
%\hat a_{A1\sigma}^\dag 
%\left[ {\mathcal R}^{x}_{B}(\theta) \right]^\dag 
%\hat a_{B1\sigma} 
%{\mathcal R}^{x}_{B}(\theta)
%e^{-i\phi\sigma} \right. \notag\\
%&&\hspace{2cm}+  \left.
%\left[ {\mathcal R}^{x}_{B}(\theta) \right]^\dag
%\hat a_{B1\sigma}^\dag 
%{\mathcal R}^{x}_{B}(\theta) 
%\hat a_{A1\sigma} e^{i\phi\sigma} \right)P_0\\&=&
\\&&\hspace*{.5cm}
=
-t_g\sum_{\sigma} P_0
\left[ 
	\hat a_{A1\sigma}^\dag 
	\left( 
		\hat a_{B1\sigma} e^{i(\phi+\psi)\sigma} \cos\frac{\theta}2 
	\right. 
\right. 
\notag \\ && \notag\hspace{3.7cm}
\left.
	- i\hat a_{B1\overline\sigma} e^{-i(\phi-\psi)\sigma} \sin\frac{\theta}2
\right)
 \\ \notag
&&\hspace{2.7cm}+   
\left( 
	\hat a_{B1\sigma}^\dag e^{-i(\phi+\psi)\sigma}  \cos\frac{\theta}2 
\right.
\notag\\ && \hspace{3.2cm}
\left.
	\left.
		+ i\hat a_{B1\overline\sigma}^\dag e^{i(\phi-\psi)\sigma} \sin\frac{\theta}2 
	\right)
	\hat a_{A1\sigma} 
\right]P_0. \notag
\end{eqnarray}
%\begin{widetext}
The generalization of the  matrix elements in Eqs. (\ref{eqn:MEt}) for this rotated operator is trivial.
Further, it is straightforward to confirm that the spin operators transform as expected under rotation when written in terms of Eq. (\ref{eqn:trans}). %Explicitly,
Therefore 
\begin{eqnarray}
&&\hspace*{-0.5cm}\left[ {\mathcal R}^{(1/2)}_{B}(-{\bm\vartheta})\right]^\dag H_{J_g} {\mathcal R}^{(1/2)}_{B}(-{\bm\vartheta}) \notag\\
&&=
% J_cP_0\Bigg[  S_{A1}^z 
%\left[ {\mathcal R}^{x}_{B}(\theta) \right]^\dag
%\left[ {\mathcal R}^{z}_{B}(\phi) \right]^\dag 
% S_{B1}^z
%{\mathcal R}^{z}_{B}(\phi) {\mathcal R}^{x}_{B}(\theta)  
%\notag\\&& \hspace{1.cm} 
%+\frac12\bigg( S_{A1}^+ 
%\left[ {\mathcal R}^{x}_{B}(\theta) \right]^\dag
%\left[ {\mathcal R}^{z}_{B}(\phi) \right]^\dag 
% S_{B1}^-
%{\mathcal R}^{z}_{B}(\phi) {\mathcal R}^{x}_{B}(\theta)  
%\notag\\&& \hspace{2cm} 
%+S_{A1}^- 
%\left[ {\mathcal R}^{x}_{B}(\theta) \right]^\dag
%\left[ {\mathcal R}^{z}_{B}(\phi) \right]^\dag 
% S_{B1}^+
%{\mathcal R}^{z}_{B}(\phi) {\mathcal R}^{x}_{B}(\theta)  \bigg)
%\notag\\&& \hspace{1.cm} 
%-\frac{\hat n_{A1}
%\left[ {\mathcal R}^{x}_{B}(\theta) \right]^\dag
%\left[ {\mathcal R}^{z}_{B}(\phi) \right]^\dag 
%\hat n_{B1}
%{\mathcal R}^{z}_{B}(\phi) {\mathcal R}^{x}_{B}(\theta) 
%}{4} \Bigg]P_0 \\
%&=&
% J_cP_0\Bigg[  S_{A1}^z 
%\left[ {\mathcal R}^{x}_{B}(\theta) \right]^\dag
% S_{B1}^z
%{\mathcal R}^{x}_{B}(\theta)  
%\notag\\&& \hspace{1.cm} 
%+\frac12\bigg( S_{A1}^+ 
%\left[ {\mathcal R}^{x}_{B}(\theta) \right]^\dag
% S_{B1}^-
% {\mathcal R}^{x}_{B}(\theta)  e^{-i\phi} 
%%\notag\\&& \hspace{2cm} 
%+S_{A1}^- 
%\left[ {\mathcal R}^{x}_{B}(\theta) \right]^\dag
% S_{B1}^+
%{\mathcal R}^{x}_{B}(\theta) e^{-i\phi}  \bigg)
%\notag\\&& \hspace{1.cm} 
%-\frac{\hat n_{A1}
%\hat n_{B1}
%}{4} \Bigg]P_0\\
%&=&
 J_cP_0\Bigg\{  S_{A1}^z 
\left[  S_{B 1}^z\cos{\theta} 
+ \left( S_{B 1}^y \cos{\psi} +  S_{B 1}^x \sin\psi \right)\sin{\theta} \right]  
\notag\\&& \hspace{1.cm} 
+\frac12\bigg[ S_{A1}^+ 
\left( S_{B1}^- \frac{1+\cos\theta}{2} e^{-i\phi} 
+  S_{B1}^+  \frac{1-\cos\theta}{2} e^{i\phi}
\right.
\notag\\&&\hspace{3.cm} 
\left.
+i\hat S_{B1}^z  \sin{\theta} \right)  e^{-i\psi} 
\notag\\&& \hspace{1.5cm} 
+S_{A1}^- 
\left( S_{B1}^+  \frac{1+\cos\theta}{2} e^{i\phi}
+  S_{B1}^- \frac{1-\cos\theta}{2} e^{-i\phi} 
\right.
\notag\\&&\hspace{3.cm} 
\left.
\left. - i\hat S_{B1}^z  \sin{\theta} \right)  e^{i\psi} \right]
\notag\\&& \hspace{1.cm} 
-\frac{\hat n_{A1}
\hat n_{B1}
}{4} \Bigg\}P_0.
\end{eqnarray}
%\end{widetext}
%Again, the generalization of the  matrix elements in Eqs. (\ref{eqn:MEJ}) for this rotated operator is straightforward - although because of the nature of the Mathematica code it is implemented via Eqs. (\ref{eqn:trans}).

However, the final effective Hamiltonian should be written in a well defined single frame. In a specific material one would usually choose this frame from crystallographic considerations. But, as we are currently considering the general case we choose to work in the local coordinate system of the $A$ molecule (see Fig. \ref{fig:lattice}{a}). Thus the pseudospin on the $B$ molecule in the effective model should be rotated back into the local frame of $A$ molecule. As we are dealing with effective spin-one degrees of freedom the appropriate transformations are 
\begin{eqnarray}
{\mathcal R}^{(1)}_{\mu}({\bm\vartheta})=\exp\left( -i{\bm\vartheta}\cdot \bm{\mathcal S}_\mu \right). \label{trans1}
\end{eqnarray}

\begin{figure}[tb]
	\begin{center}
		\includegraphics[width=0.9\columnwidth]{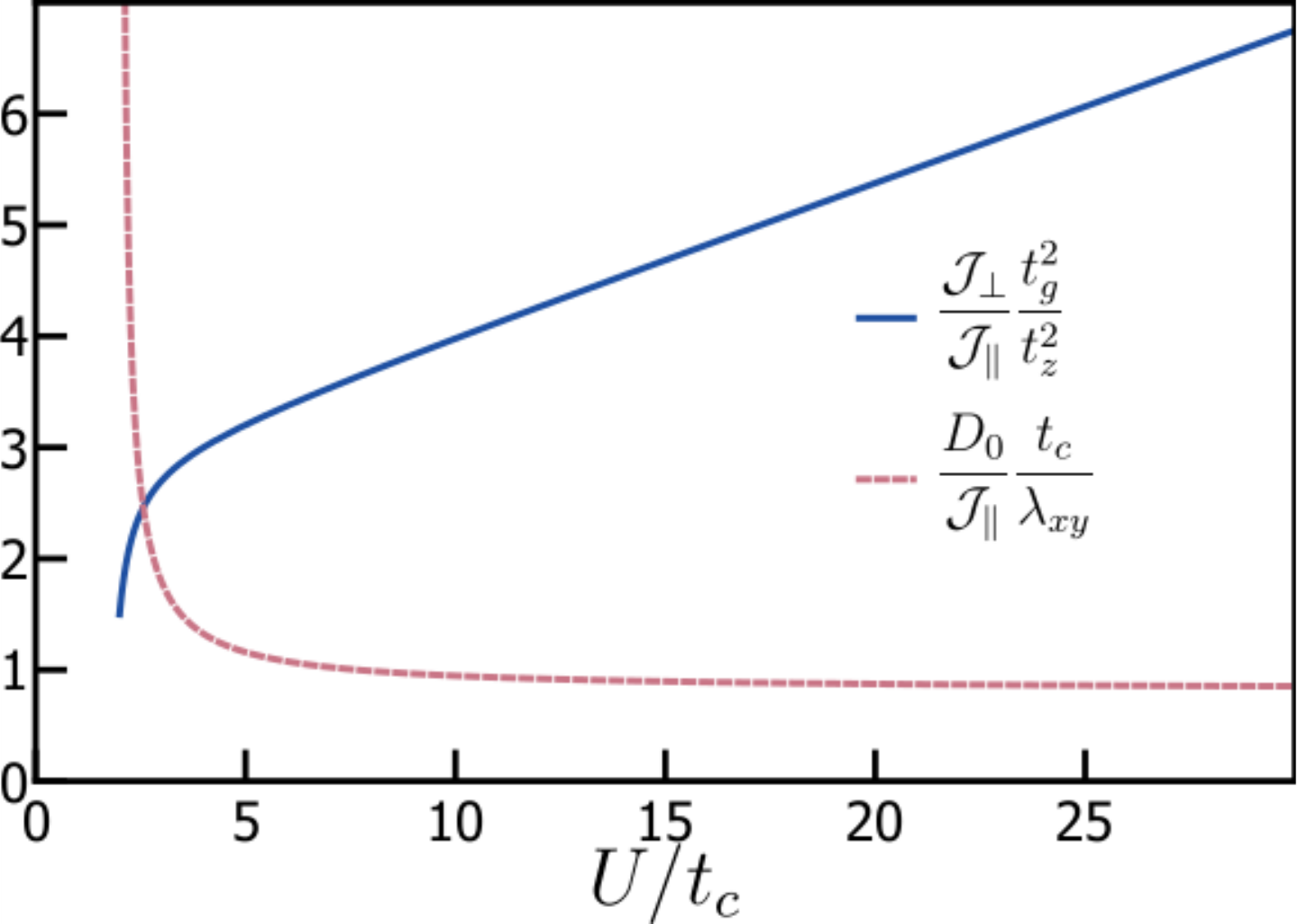}
	\end{center}
	\caption{Relative strengths of the effective  \DM and exchange interactions as a function of the interaction strength. Here we have set $J_c=4t_c^2/U$,  $J_g=4t_g^2/U$ and  $J_z=4t_z^2/U$ to reduce the parameter space. The ratio of the in-plane DM and Heisenberg exchange constants saturates to $D_0/{\mathcal J}_\|=\sqrt{2}(\lambda_{xy}/t_c)[(19/33)+(92/121)(t_c/U)+{\mathcal O}((t_c/U)^2)]$ in the strongly correlated regime ($U\rightarrow\infty$); whereas ${\mathcal J}_\perp/{\mathcal J}_\|=(t_z/t_g)^2[(3/22)(U/ t_c)+(324/121)+(702/1331)(t_c/U) + {\mathcal O}((t_c/U)^2)]$ grows linearly.} 
	\label{fig:ratios}
\end{figure}

Carrying out this process yields the effective pseudospin Hamiltonian:
\begin{eqnarray}
H_{S=1}^\text{eff}&=&-2Nt_c+\sum_{\mu\nu}\Bigg[{\mathcal J}_{\mu\nu}\left( \bm{\mathcal S}_\mu\cdot\bm{\mathcal S}_\nu -  \frac{\hat n_\mu^{(h)} \hat n_\nu^{(h)}}{4} \right) \notag\\&& \hspace*{2.5cm}
+\bm D_{\mu\nu}.\bm{\mathcal S}_\mu\times\bm{\mathcal S}_\nu\Bigg], \label{HDM}
\end{eqnarray}
where $\mathcal J_{\mu\nu}=\mathcal J_\|$ [cf. Eq. (\ref{EqJ})] and
\begin{subequations}
\begin{eqnarray}
D_{\mu\nu}^x&=&%\frac12\left(D^+-D^-\right)=
D_0  \left( 
	\sin\phi \cos\theta \cos\psi 
	+  \cos\theta \sin\psi
\right),  \label{Dx}
\\
D_{\mu\nu}^y&=&%-\frac{i}2\left(D^+-D^-\right)=
D_0 \left(
	\cos\phi \cos\theta \cos\psi
	-\sin\theta \sin\psi
	-1
\right),  \label{Dy}\\
D_{\mu\nu}^z&=&D_0  \sin\theta \cos\psi, \label{Dz} \\
D_0 &=& \sqrt{2}\frac{\lambda_{xy}}{t_c} \left[ \frac{\left(8 t_c-J_c\right)J_g}{54 \left(2 t_c-J_c\right) }
%\right.\notag\\&& \hspace*{0.75cm} \left.
- \frac{J_c(4t_c+J_c)t_g^2}{243(2t_c-J_c)^2t_c} \right]. \hspace*{0.8cm} \label{EqD0}
\end{eqnarray}
\label{EqD}
\end{subequations}
%where we have used the fact the $J_g/J_c=(t_g/t_c)^2$ of the rhs of several eqns to eliminate $t_g$. (Is this helpful?)
Note {that} $\lambda_z$ does not appear -- this can be readily understood as the DM coupling arises from the transfer of angular momentum between the spin and orbital degrees of freedom and the molecular orbitals of a $C_3$ symmetric molecule only carry angular momentum around the $z$-axis.  Fig. \ref{fig:ratios} displays the variation of $D_0/{\mathcal J}_\|$  with the strength of the electronic correlations. One clearly sees that this ratio saturates in the strongly correlated limit.

%\subsection{Dzyaloshinskii-Moriya interaction}
%
%\begin{eqnarray}
%{\mathcal H}_{DM}&=&{\bm D}.{\bm S}_1\times{\bm S}_2\\
%&=&D^x(S_1^y S_2^z - S_1^z S_2^y)
%+D^y(S_1^z S_2^x - S_1^x S_2^z)
%+D^z(S_1^x S_2^y - S_1^y S_2^z) \label{DMxyz} \\
%&=&\frac{i}2\left[D^+(S_1^- S_2^z - S_1^z S_2^-)
%+D^-(S_1^z S_2^+ - S_1^+ S_2^z)
%+D^z(S_1^+ S_2^- - S_1^- S_2^+) \label{DMpm} \right]
%\end{eqnarray}
%where $D^\pm=D^x\pm iD^y$. Note that Eq. (\ref{DMpm}) follows regardless of the magnitude of $S_1$ or $S_2$. Therefore it is clear that the $\bm D=(D^x,D^y,D^z)$ given by Eqs. (\ref{Dm}-\ref{Dz}) is exactly the Dzyaloshinskii-Moriya interaction.

%\subsection{Effective Hamiltonian}
%
%Thus we have,
%\begin{eqnarray}
%H_{S=1}^\text{eff}=\varepsilon_0+J\bm S_A\cdot\bm S_B+\bm D.\bm S_A\times\bm S_B
%\end{eqnarray}

\subsubsection{Molecules related by a $\pi$-rotation about the $z$-axes ($C_2^z$ symmetry)}\label{sect:rot}

We now consider a pair of molecules related by a rotation of $\pi$ about a z-axis{, Fig. \ref{fig:lattice}d}. In the absence of SMOC the tight-binding model is identical to that of a pair of molecules related by an inversion center. But $C_3$ molecules have structures that differentiate between ``up" and ``down" physical orientations, e.g., the $\mu_3$ sulphur in \Mo (above the center of the three Mo atoms, see Fig. \ref{fig:Wannier}), has no counterpart below the plane of the molecule. {Thus it is clear that the \Mo molecules in the plane are related by an inversion center and not a rotation. Nevertheless, in other materials the converse may be the case and so we briefly consider molecules related by a $\pi$ rotation in this section.}

 $C_2^z$ and inversion have different effects on $H_{SMO}^{(\mu)}$. 
Inversion leaves $H_{SMO}^{(\mu)}$ unchanged as both ${\bm L}_\mu$ and ${\bm S}_\mu$ are pseudovectors. However,  phases arise under $C_2^z$ rotations. Spin-$\frac12$ particles pick up a phase of $\pi\sigma_z$, cf. Eq. (\ref{eqn:trans}), while angular momentum-1 particles pick up a phase of $-\pi L^z_\mu$, cf. Eq. (\ref{trans1}). Therefore, $\lambda_z^{(\mu)}=\lambda_z$ on all molecules, but $\lambda_{xy}^{(\mu)}=\lambda_{xy}=-\lambda_{xy}^{(\nu)}$,  where $\mu$ and $\nu$ are a pair of molecules related by a $C_2^z$ symmetry. 

Carrying out the perturbation theory as in the inversion symmetric case one finds that the effective low-energy model is described by Hamiltonian (\ref{HDM}) with ${\mathcal J}_{\mu\nu}={\mathcal J}_\|$ [cf. Eq. (\ref{EqJ})]  and the Dzyaloshinskii-Moriya interaction given by $\bm{D}_{\mu\nu}=(0,D_0,0)$, where $D_0$ is given by Eq. (\ref{EqD0}).
%This result is unique to the molecular case. 
We stress that it is the  internal structure  of the molecule that drives the differences between the inversion and $C_2^z$ symmetric case.

\subsubsection{Broken $C_2^z$ symmetry} \label{sect:bC2}

We can consider rotating one of the pair of molecules without affecting the intra-molecular terms in the Hamiltonian. This is entirely analogous to the calculation in section \ref{sect:brokenI}. Again we find that the effective low-energy model is described by Hamiltonian (\ref{HDM}), the parameters of the model are as in Eq. (\ref{EqJ}) and (\ref{EqD}) except that
\begin{eqnarray}
D^y_{\mu\nu}=D_0 \left(
\cos\phi \cos\theta \cos\psi
-\sin\theta \sin\psi
+1
\right).
\end{eqnarray}

\subsection{Interlayer coupling}\label{sect:interlayer}

We now consider a pair of molecules coupled by three hopping integrals of equal strength, $t_z$, such that electrons in the $i$th Wannier orbital on molecule $\mu$ can hop to the $i$th Wannier on molecule $\nu$, cf. Fig \ref{fig:lattice}b.
To allow for a consistent treatment of the Hubbard model we also include a superexchange interaction between the $i$th Wanniers on molecule $\mu$ and $\nu$; $J_z=4t_z^2/U+{\mathcal O}(t_z^4/U^4)+{\mathcal O}(t_z^2t_c^2/U^4)$. Thus the intermolecular coupling Hamiltonian is
\begin{eqnarray}
H_z &=&  P_0\sum_{\mu\nu}\left[\sum_{\sigma,j=1}^{3}t_z\left(\hat{{h}}^\dagger_{\mu j\sigma}\hat{{h}}_{\nu j\sigma}+H.c.%\hat{a}^\dagger_{Bj\sigma}\hat{a}_{Aj\sigma}
\right)
\right. \notag\\
&& \hspace{0.8cm}
+\left. J_z\left( \bm S_{\mu j}\cdot\bm S_{\nu j} -  \frac{\hat n_{\mu j}^{(h)} \hat n_{\nu j}^{(h)}}{4} \right)\right]P_0 \label{HamZ}
\end{eqnarray} 
This describes the dominant interlayer coupling in \Mo \cite{JackoPRB15}. If a pair of molecules, $\mu$ and $\nu$, are related by translational symmetry [as is the case for layers of Mo$_3$S$_7$(dmit)$_3$] $H_{SMO}^{(\mu)}=H_{SMO}^{(\nu)}$.

To first order in $H_{SMO}$ and $J_z$ and second order in $t_z$ we find that there is no DM coupling. This is not a consequence of symmetry and indeed we will see below that an interlayer DM interaction is induced by longer range SOC.
Thus, the effective Hamiltonian is that given in Eq. (\ref{HnoDM}), but with the effective Heisenberg exchange constant ${\mathcal J}_{\mu\nu}={\mathcal J}_\perp$  for nearest neighbors $\mu$ and $\nu$ perpendicular to the plane [i.e., with the intermolecular coupling  between each pair of equivalent orbitals, as described by Eq. (\ref{HamZ}), cf. Fig. \ref{fig:lattice}b] where
	\begin{eqnarray}
	\label{EqJz}
	{\mathcal J}_\perp=\frac{J_z}{3}  
	+\frac{4t_z^2}{ 9(2t_c-J_c)}.
	\label{JperpZ}
	\end{eqnarray}
	
It is interesting to note, cf. Fig. \ref{fig:ratios}, that ${\mathcal J}_\|\rightarrow0$ as $J_c\rightarrow0$ ($U\rightarrow\infty$), cf. Eq. (\ref{EqJ}). However, ${\mathcal J}_\perp$ does not vanish in that limit. We have previously observed this numerically in the Hubbard model of \Mo \cite{Jaime16}. The Hubbard model results asymptote towards the \tJ results as $U\rightarrow\infty$, but only very slowly. Nevertheless, the current analytical treatment allows us to gain a deeper understanding of  this emergent quasi-on{e}-dimensionality.

\begin{figure}[tb]
	\begin{center}
		\includegraphics[width=0.9\columnwidth]{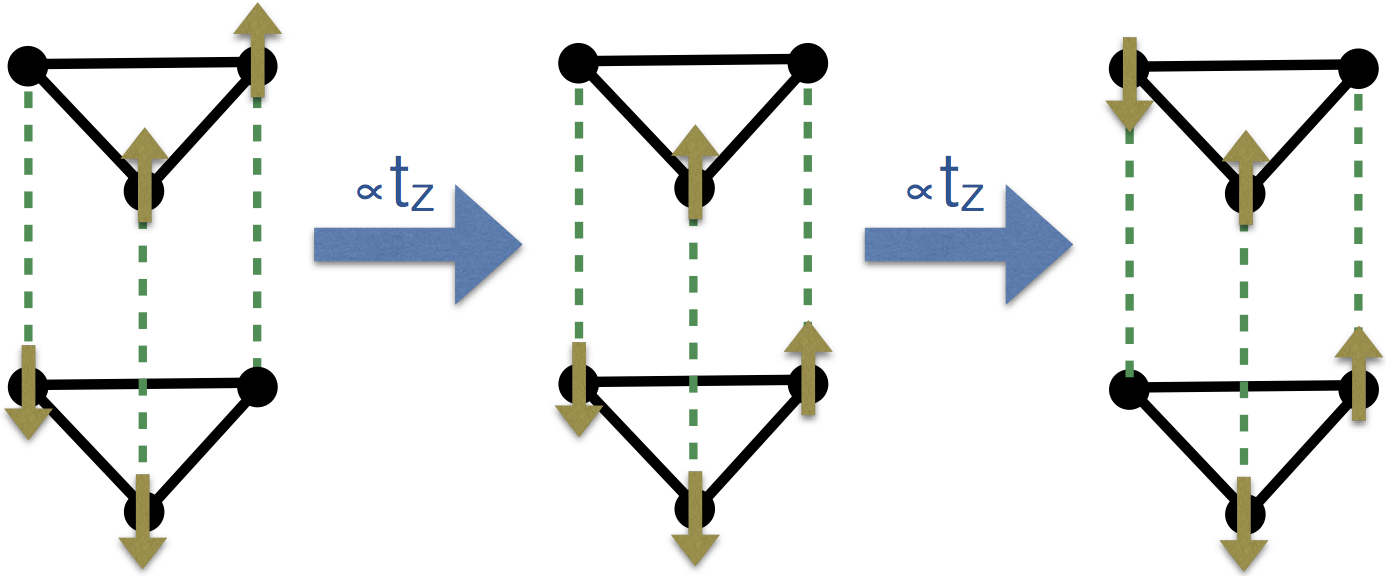}
		{%\normalsize 
			\large $\Delta E\propto t_c$}%[1+\cos(2\pi k_1/3)]$}
	\end{center}
	\caption{Classical cartoon of a process that contributes to ${\mathcal J}_\perp$ in the strongly correlated limit, $J_c, J_z\rightarrow0$ ($U\rightarrow\infty$).  On the far left we sketch of one of the states in the low-energy subspace, $\ket{{\mathcal S}^z_\text{upper}=1,{\mathcal S}^z_\text{lower}=-1}$, with holes marked by  (yellow) arrows; the actual single molecule eigenstates are linear superpositions of cyclic permutations of the states sketched here. One hole hops from the upper molecule to the lower molecule along the right hand bond with an amplitude proportional to $t_z$. This leaves an intermediate state that is higher energy in energy than the initial state by $\Delta E\equiv E_5^{2\Gamma'}(k_1,\sigma)+E_3^{(2S+1)\Gamma}-2E_4$. For $\lambda_z=\lambda_{xy}=0$ and $J_c\rightarrow0$ we have $E_3^{(2S+1)\Gamma}=0$ for all $S, \Gamma$ and $E_5^{2\Gamma'}(k_1,\sigma)=2t_c\cos(2\pi k_1/3)$ for all $\Gamma', \sigma$; hence $\Delta E=2t_c[1+\cos(2\pi k_1/3)]$. This intermediate excited state has an amplitude proportional to $t_z$ for the hole on the left hand site to hop back to the top molecule giving the state $\ket{0,0}$, which is again part of the low-energy manifold. From this, and similar, processes one expects ${\mathcal J}_\perp\propto t_z^2/t_c$ in the strongly correlated limit, as we find explicitly, cf. Eq. (\ref{JperpZ}). However, this classical picture is somewhat oversimplified because the holes are not localized in the true single molecule eigenstates. Therefore a more complete treatment must include quantum mechanical interference, as described in the main text (section \ref{sect:interlayer}). 
	} 
	\label{fig:why}
\end{figure}

\begin{figure}[tb]
	\begin{center}
		\includegraphics[width=0.9\columnwidth]{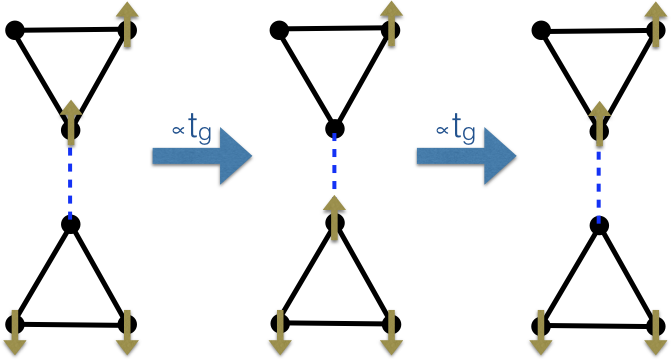}
		{%\normalsize 
			\large $\Delta E\propto t_c$}
	\end{center}
	\caption{
		{
			Classical cartoon illustrating the suppression of ${\mathcal J}_\|\rightarrow0$ in the strongly correlated limit, $J_c, J_g\rightarrow0$ ($U\rightarrow\infty$). Electrons can still hop between the two molecules, but no processes at second order can change the net spin on either molecule. Thus, there is a constant offset in the energy at second order, but the effective exchange coupling between molecules vanishes.  
		} 
	} 
	\label{fig:why_not}
\end{figure}

Consider the {processes sketched in Figs. \ref{fig:why} and \ref{fig:why_not}.} 
A hole hops between a pair of molecules along a particular `bond'. One molecule now contains one hole -- the states of this eigenstates  on this molecule are  trivial Bloch states (see Appendix \ref{append5e}), and will not concern us further. The other molecule contains three holes and is described by a Heisenberg model in the $U\rightarrow\infty$ limit. The eigenstates (cf. Appendix \ref{append3e}) are a two spin-doublets with energy $E_3^{2E}=-3J_c/2$ and a spin-quadruplet with energy $E_3^{4A}=0$. Note that as $J_c\rightarrow0$ these states become (eight-fold) degenerate. This enhances the effects of interference  between the different intermediate excited states in the sum in Eq. (\ref{eqn:MEt}). The interference is destructive if the system returns to the low-energy subspace by a hole hopping along the same `bond' as in the initial step; but constructive if the second hop is along a different bond. This explains why ${\mathcal J}_\|$ vanishes in this limit, but ${\mathcal J}_\perp$ does not. Conversely, if $J_c$ is increased from zero the degeneracy between the spin-quadruplet and the spin-doublets is lifted and the interference is suppressed.

\section{Spin-orbit coupling on the intermolecular bonds}\label{sect:2mol-SOI}

So far we have assumed that SOC is a purely intramolecular effect. This is not  true in general. The Pauli equation  \cite{Dyall} gives the SOC as 
\begin{eqnarray}
H_\textrm{SO}^\textrm{Pauli}&=&\frac{\hbar}{4m^2c^2}(\bm p\times\bm\nabla V(\bm{r}))\cdot\bm \sigma%\notag\\&
%\equiv%&
%\bm K\cdot\bm \sigma
,
\label{Pauli}
\end{eqnarray}
where $\bm p$ is the electronic momentum operator and $V(\bm{r})$ is the potential in which the electrons move. It is straightforward to calculate the matrix elements between Wannier spin-orbitals $\ket{j,\alpha}$ for this interaction which yields
\begin{eqnarray}
\bra{\mu,i,\alpha} H_\textrm{SO}^\textrm{Pauli} \ket{\nu,j,\beta}&=&i{\bm \lambda}_{\mu\nu;ij} \cdot \bm \sigma_{\alpha\beta},
\label{su2}
\end{eqnarray}
where the pseudovectors ${\bm \lambda}_{\mu\nu;ij}$ are real, material specific, constants; the reality of the ${\bm \lambda}_{\mu\nu;ij}$ is a consequence of SU(2) invariance.

Carrying out the perturbation theory to first order in 
\begin{eqnarray}
H_\textrm{SOI}\equiv i\sum_{\mu\nu}\sum_{i,j=1}^3 \sum_{\alpha\beta} {\bm \lambda}_{\mu\nu;ij}  \cdot \bm \sigma_{\alpha\beta} \hat{a}^\dagger_{\mu i\alpha}\hat{a}_{\nu j\beta}+H.c.
\end{eqnarray}
and $H_{SMO}$ and second order in 
\begin{eqnarray}
H_t &=&  P_0\sum_{\mu\nu\sigma}\sum_{i,j=1}^{3}t_{\mu\nu;ij}\left(\hat{a}^\dagger_{\mu i\sigma}\hat{a}_{\nu j\sigma}+H.c.%\hat{a}^\dagger_{Bj\sigma}\hat{a}_{Aj\sigma}
\right)P_0
\end{eqnarray} 
[the intramolecular exchange terms do not contribute to the DM interaction beyond terms similar to Eq. (\ref{EqD0}), and so we neglect these for simplicity] one finds that
\begin{subequations}
	\begin{eqnarray}
	D^z_{\mu\nu} &=& \sum_{i,j=1}^3 \Upsilon_{ij}\lambda_{\mu\nu;ij}^z, \\
	D^y_{\mu\nu} &=& \Xi^y \lambda_{xy} + \sum_{i,j=1}^3  \Upsilon_{ij}\lambda_{\mu\nu;ij}^y,  \label{Dygen} \\
	D^x_{\mu\nu} &=&  \Xi^x \lambda_{xy} + \sum_{i,j=1}^3  \Upsilon_{ij}\lambda_{\mu\nu;ij}^x.  \label{Dxgen}
	\end{eqnarray}
	\label{Dgen}
\end{subequations}
The general expressions for $\Upsilon_{ij}$ and the $\Xi^\eta$ are given in Appendix \ref{App:Xi}. However, given the complexity of these expressions, it is more instructive to examine some special cases.

\subsection{Simple tube}

Here we consider the natural extension of the  interlayer model discussed in section \ref{sect:interlayer}; i.e., $t_{\mu\nu;ij}=t_{\mu\nu}\delta_{ij}$ and
\begin{eqnarray}
\hspace*{-0.3cm}
{\bm \lambda}_{\mu\nu;ij}=&\hspace*{-5pt}\Big(&\hspace*{-5pt}
\lambda_1\cos\phi_j-\lambda_2\sin\phi_j,~\notag\\&&\hspace*{-5pt}
\lambda_2\cos\phi_j+\lambda_1\sin\phi_j,~
\lambda_3
\Big)\delta_{ij},\label{Eq:C3soc}
\end{eqnarray} 
where  $\lambda_1, \lambda_2$, and $\lambda_3$ are constants. Importantly, this form of the intermolecular spin-orbit coupling correctly accounts for the mixing of the $x$ and $y$ components of ${\bm \lambda}_{\mu\nu;ij}$ under the three-fold rotations about the $z$ axis.
Thus, $\sum_{i,j=1}^{3}\lambda_{\mu\nu;ij}^x=\sum_{i,j=1}^{3}\lambda_{\mu\nu;ij}^y=0$. It then follows immediately from Eqs. (27) that ${D}_{\mu\nu}^x={D}_{\mu\nu}^y=0$. Further one finds that $\Xi^x=\Xi^y=0$ and $\Upsilon_{ij}=\delta_{ij}8 t_{\mu\nu}/[9 (J_c - 2 t_c)]$. Hence
\begin{eqnarray}
{\bm D}_{\mu\nu}={\bm {\hat z}}\frac{8t_{\mu\nu}\lambda_3}{9(J_c-2t_c)},
\end{eqnarray}
where ${\bm {\hat z}}=(0,0,1)$.

\subsection{\Mo}

Recently Wannier orbitals for \Mo have been constructed from density functional calculations \cite{JackoPRB15}. These have been used to provide parameterizations of the single particle electronic structure in terms of the tight-binding model. %Several parameterizations were provided, representing different trade-offs between simplicity and accurately reproducing the full density functional calculation. \cite{JackoPRB15} 
Furthermore, similar calculations have been reported from four-component relativistic density functional  theory \cite{Jacko16}. The hopping integrals ($t_{\mu\nu;ij}$) do not show significant changes from the initial (scalar relativistic) calculations, but this process does allow for the calculation of the parameters in $H_{SMO}$ and $H_{SOI}$ from first principles. 
This parameterization contains all of the terms in the simple in the simple tube model described above and additional terms, for example hopping and SOC between non-equivalent sites of different molecules [i.e., $t_{\mu\nu;ij}\ne t_{\mu\nu}\delta_{ij}$ and ${\bm\lambda}_{\mu\nu;ij}$ given by Eq. (\ref{Eq:C3soc})].

\begin{figure}[tb]
	\begin{center}
		\includegraphics[width=0.9\columnwidth]{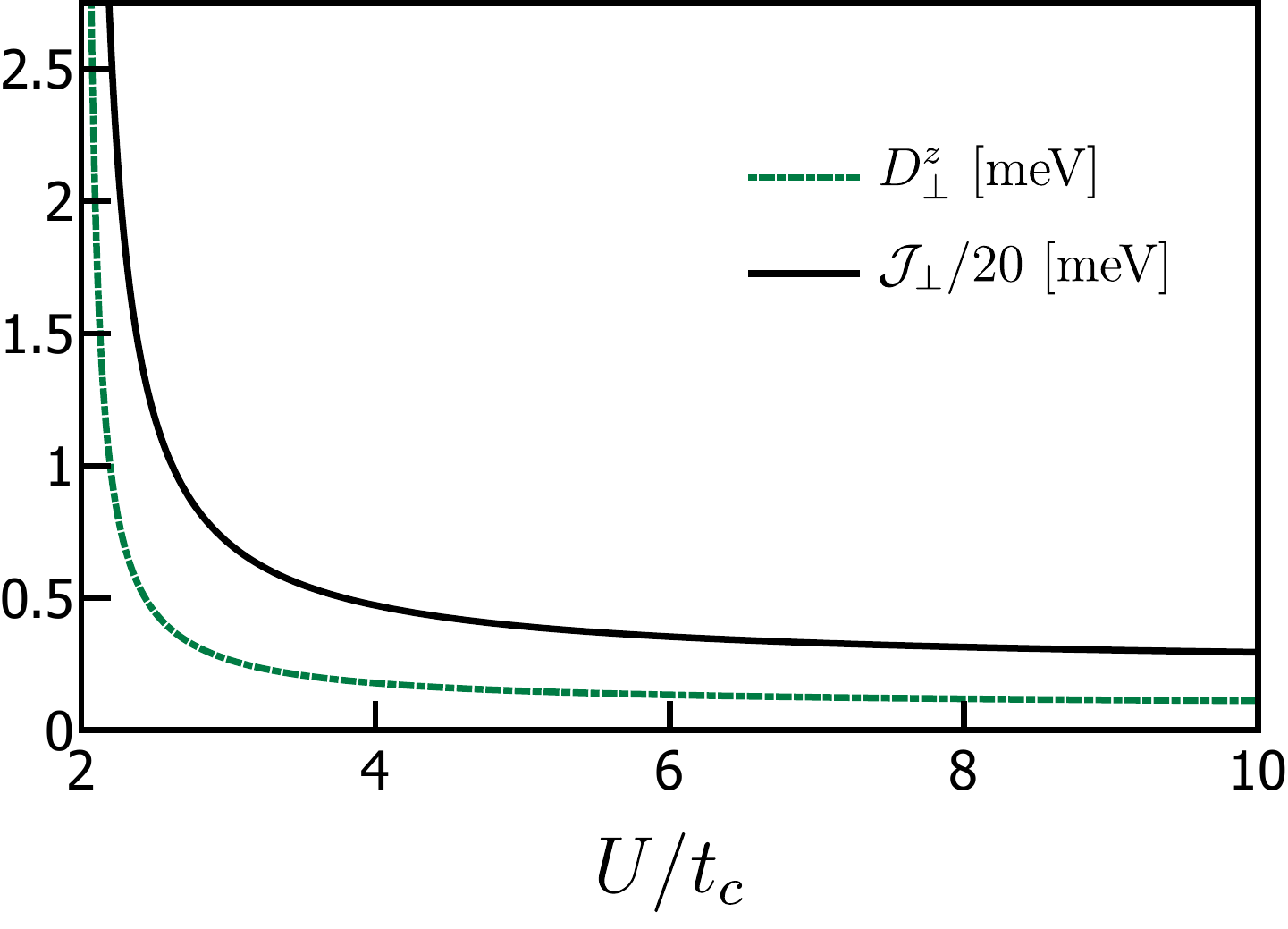}
	\end{center}
	\caption{Effective interlayer \DM and exchange interactions in \Mo as a function of the interaction strength. Here we have set $J_c=4t_c^2/U$ and taken all other parameters from first principles calculations \cite{Jacko16,JackoPRB15}. Note that the perturbation theory breaks down as $U\rightarrow2t_c$.} 
	\label{fig:DJ}
\end{figure}

First we consider nearest neighbors along the $c$-axis. 
The molecules stack above/below one another so as to retain the $C_3$ symmetry of the individual molecules. Therefore the parameters are rotationally invariant and we can write $t_{\mu\nu;ij}=t(i-j)$ and 
\begin{eqnarray}
{\bm \lambda}_{\mu\nu;ij}=&\hspace*{-5pt}\Big(&\hspace*{-5pt}
\lambda_1(i-j)\cos\phi_j-\lambda_2(i-j)\sin\phi_j,~\notag\\&&\hspace*{-5pt}
\lambda_2(i-j)\cos\phi_j+\lambda_1(i-j)\sin\phi_j,~\notag\\&&\hspace*{-5pt}
\lambda_3(i-j)
\Big),
\end{eqnarray}
where now the functions $\lambda_1(i-j), \lambda_2(i-j)$, and $\lambda_3(i-j)$ only depend on the difference of $i$ and $j$ modulo three and we have suppressed the molecular labels on the right hand sides for clarity; in both cases the subtraction is defined modulo three.  Again one has ${D}_{\perp}^x={D}_{\perp}^y=0$, indeed this  could have been anticipated from the $C_3$ symmetry of the problem \cite{Moriya}.  One finds that $\Xi^x=\Xi^y=0$ and hence the DM interaction between nearest neighbors  along the $c$-axis, ${\bm D}_\perp$, is
\begin{eqnarray}
{\bm D}_{\perp}={\bm {\hat z}}\frac{8}{9(J_c-2t_c)}\sum_{ij}t(i)\lambda_3(j). \label{Dperp}
\end{eqnarray}

Similarly, one finds that the effective exchange interaction between nearest neighbors in  the $c$-direction, ${\mathcal J}_\perp$ is
\begin{eqnarray}
{\mathcal J}_\perp=\frac{8}{9 (2 t_c-J_c)}
\sum_{ij} t(i) t(j) \left(\delta_{ij}-\frac12\right). \label{Jperp}
\end{eqnarray}
We plot the variation of ${\bm D}_\perp$ and ${\mathcal J}_\perp$ in Fig. \ref{fig:DJ}, where we have parameterized $J_c=4t_c^2/U$. Note that the perturbation theory breaks down for large $J_c$ (small $U$).

It is interesting to note that the common denominators of Eqs. (\ref{Dperp}) and (\ref{Jperp}) imply that the ${{\bm D}_\perp}/{{\mathcal J}_\perp}$ is independent of $J_c$, which is not accurately known in \Mo {due to the absence of a first principles calculation of $U$}. For the tight-binding and SOC parameters calculated from first principles for \Mo one finds that ${{\bm D}_\perp}/{{\mathcal J}_\perp}=0.019{\bm {\hat z}}$. This could be enhanced, by an order of magnitude or more, by moving to systems where the molybdenum is substituted by tungsten and/or the sulfur is substituted by selenium \cite{Amie16}; suggesting that in such materials the \DM interaction will play a significant role.

Nearest neighbors in the basal plane of \Mo are separated by an inversion center. It therefore follows trivially that the in-plane \DM interaction vanishes \cite{Moriya}. Thus in the large{-}$U$ limit the effective Hamiltonian is that of a spin-one Heisenberg chain with DM interaction:
\begin{eqnarray}
H^\text{eff}_\text{Mo}=\sum_\mu \left[ {\mathcal J}_\perp\bm{\mathcal S}_\mu\cdot\bm{\mathcal S}_{\mu+1}
+   {\bm D}_\perp \cdot \bm{\mathcal S}_\mu\times\bm{\mathcal S}_{\mu+1} \right] ,
\end{eqnarray} 
where, we have dropped constant terms. Without loss of generality we can choose the quantization ($z$) axis for the effective spin ($\bm{\mathcal S}_\mu$) to lie parallel to ${\bm D}_\perp$, which points along the crystallographic $c$-axis, yielding
\begin{eqnarray}
H^\text{eff}_\text{Mo}&=&\sum_\mu\bigg\lbrace {\frac12} \left[ ({\mathcal J}_\perp + iD_\perp) {\mathcal S}_\mu^+\cdot{\mathcal S}_{\mu+1}^- +H.c.\right] \notag\\&&\hspace*{1cm}
+ {\mathcal J}_\perp {\mathcal S}_\mu^z\cdot{\mathcal S}_{\mu+1}^z \bigg\rbrace,
\end{eqnarray} 
where $D_\perp=|{\bm D}_\perp|$. The gauge transformation ${\mathcal S}_{\mu}^- \rightarrow {\mathcal S}_{\mu}^-\exp[\mu\tan^{-1}(D_\perp/{\mathcal J}_\perp)]$ gives
\begin{eqnarray}
H^\text{eff}_\text{Mo}&=&{\mathcal J}^*\sum_\mu \left[  {\mathcal S}_\mu^x\cdot{\mathcal S}_{\mu+1}^x + {\mathcal S}_\mu^y\cdot{\mathcal S}_{\mu+1}^y 
+ \Delta^* {\mathcal S}_\mu^z\cdot{\mathcal S}_{\mu+1}^z \right],\notag\\
\end{eqnarray} 
where ${\mathcal J}^*=\sqrt{{\mathcal J}_\perp^2 + D_\perp^2}$ and $\Delta^*=\cos(D_\perp/{\mathcal J}_\perp)$. For the pure spin-one model the Haldane phase is believed to be stable for $0 \lesssim \Delta^*/{\mathcal J}^* \lesssim 1.2$ \cite{Su}.

The large charge fluctuations expected for \Mo \cite{Henry} imply that the value of the Haldane gap is likely to be strongly suppressed from the value of the spin-one Heisenberg chain \cite{JananiPRB} (where the gap $\sim0.4{\mathcal J}_\perp$ \cite{ref10}).
Nevertheless $D_\perp=0.02{\mathcal J}_\perp$ is still likely small compared to the Haldane gap and so the DM interaction is unlikely to destabilize the Haldane phase. However, this does not mean that it does not have important physical effects. For example, DM coupling is known to lead to a significant enhancements in the magnetic susceptibility \cite{Huang}; furthermore it has been argued that this is relevant to Ni(C$_2$H$_8$N$_2$)$_2$NO$_2$(ClO$_4$) where the DM coupling is estimated to be the same order of magnitude as we have calculated above \cite{Huang}. It has also been proposed that the DM interaction will lead to significant changes in the electron spin resonance spectrum \cite{Karimi}.

\section{Conclusions}

We found that the effective Heisenberg exchange coupling constants, ${\mathcal J}_{\mu\nu}$, are strongly dependent on how the (spatially separated, cf. Fig. \ref{fig:Wannier}) Wannier orbitals on the two molecules couple. If the intramolecular coupling is purely through a single orbital on each molecule then ${\mathcal J}\rightarrow0$ in the strongly correlated limit. In contrast if all three Wannier orbitals couple to the equivalent orbital on the neighboring molecule ${\mathcal J}\ne0$ in the strongly correlated limit. This can be understood by considering the interference between the many different intermediate excited states contributing to effective exchange. In the former case the interference is purely destructive, whereas the latter case also contains constructive interference effects, which allows the effect exchange  coupling to remain non-zero even as $U\rightarrow\infty$.

We have also seen that in multi-nuclear coordination complexes the DM interaction is strongly dependent on factors that are absent or significantly different in the atomic crystals (as opposed to molecular crystals), such as transition metal oxides.

We found that the nature of the coupling between molecules, i.e., which orbitals electrons can hop between and the relative strength of this hopping and its spin-orbit coupled analogue  strongly affects  the nature of the DM interaction. This  effect is somewhat analogous to atomic systems where multiple orbitals on a single atom are relevant {\cite{Yildrum,perkins2014a,Perkins}}.

If two spherically symmetric objects are brought together, the new system is inversion symmetric and therefore cannot have a DM interaction. In atomic crystals the spherical symmetry of the atom is broken by the crystal field due to its local environment. It is the relative orientation of the local environments of the heavy atoms with unpaired electrons that determine{s} the nature of the DM interaction. In molecular crystals the fundamental building block is not spherically symmetric. Therefore, one does not require a strong crystal field to observe a DM interaction.

Furthermore, a wide variety of molecular packing motifs and angles are found in the vast array of molecular crystals. In contrast, the chemistry of, say, transition metal oxides means that the vast majority of materials  have similar structures -- typically $90^\circ$ or $180^\circ$ metal-oxygen-metal angles and  distortions of these structures. Therefore, one might expect a greater range of possibilities to be realized in molecular crystals. However, quantum interference due to hopping between Wannier orbitals on the same molecule  leads to new effects not seen in atomic crystals where the atomic orbitals on any individual atom can always be chosen {so that there is no direct hopping between them}.

On the basis of the above results and recent DFT calculations of the SOC in \Mo we argued that the Haldane phase  is likely to be stable to the DM interaction  in \Mo, despite the strong charge fluctuations previously predicted due to the internal electron dynamics of within the molecules.  Nevertheless, comparison with previous calculations suggest that the DM interaction may lead to experimentally observable changes in \Mo, such as an enhancement of the magnetic susceptibility.

{
	Key experimental tests of these predictions include the detection of the spin-1/2 edge states characteristic of the Haldane phase \cite{AKLTlong} in \Mo. Suitable probes include electron spin resonance (ESR) \cite{Hagiwara} or nuclear magnetic resonace (NMR) \cite{Tedoldi}. Magnetic resonance experiments should also be sensative to the enhancement of the magnetic susceptibility because of the DM coupling \cite{Huang,Karimi}. Furthermore, the chemical replacement of \Mo with non-magnetic impurities should lead to dramatic changes density of edge spins. 
}

{
	A number of synthetic approaches are also suggested by this work. For example, growing materials with heavier metals, or replacing S by Se or Te should singificantly enhance the SMOC \cite{Amie16}.
	More exotically, monolayer films could open up the possibility of controlling the interactions of within a two-dimensional material. 
}

\section*{Acknowledgments}

This work was supported by the Australian Research Council through Grants No. FT130100161, DP130100757 and DP160100060. J.M. acknowledges financial support from (MAT2015-66128-R) MINECO/FEDER, UE.

\appendix

\begin{widetext}

\section{Excited states of the neutral molecule}   \label{sect:neut}

%	\begin{subequations}
		
		The lowest lying excited state of the neutral molecule is a doublet with energy
		$E_4^{1E}=-(J_c/2) - t_c$ to first order in the SMOC. The wavefunctions are
		\begin{eqnarray}
		\ket{\Phi_4^{1E}(1)}	&=& \frac{1}{\sqrt{6}}\sum_{j} \left[ 
		e^{i\phi_{j+1}}\left( \hat{a}_{\mu j\uparrow} \hat{a}_{\mu j-1\downarrow} - \hat{a}_{\mu j\downarrow} \hat{a}_{\mu j-1\uparrow}  \right) 
		+\frac{i\lambda_{xy}}{\sqrt{3} (J_c - 2 t_c)}  \hat{a}_{\mu j\uparrow} \hat{a}_{\mu j-1\uparrow} 
		+\frac{2i\lambda_{xy}e^{-i\phi_{j+1}}}{\sqrt{3} (J_c +4 t_c)}   \hat{a}_{\mu j\downarrow} \hat{a}_{\mu j-1\downarrow} 
		\right. \notag\\&& \hspace{1cm} \left.
		+\frac{i\lambda_z  e^{i\phi_{j+1}}}{\sqrt{3} (J_c +4 t_c)} \left( \hat{a}_{\mu j\uparrow} \hat{a}_{\mu j-1\downarrow} + \hat{a}_{\mu j\downarrow} \hat{a}_{\mu j-1\uparrow}  \right) 
		\right] \ket{vac_h}, 
		\\
		\ket{\Phi_4^{1E}(-1)}	&=& \frac{1}{\sqrt{6}}\sum_{j} \left[ 
		e^{-i\phi_{j+1}}\left( \hat{a}_{\mu j\uparrow} \hat{a}_{\mu j-1\downarrow} - \hat{a}_{\mu j\downarrow} \hat{a}_{\mu j-1\uparrow}  \right) 
		-\frac{i\lambda_{xy}}{\sqrt{3} (J_c - 2 t_c)} \hat{a}_{\mu j\downarrow} \hat{a}_{\mu j-1\downarrow} 
		\right. \notag\\&& \hspace{1cm} \left.
		+\frac{i\lambda_z  e^{-i\phi_{j+1}}}{\sqrt{3} (J_c +4 t_c)} \left( \hat{a}_{\mu j\uparrow} \hat{a}_{\mu j-1\downarrow} + \hat{a}_{\mu j\downarrow} \hat{a}_{\mu j-1\uparrow}  \right) 
		-\frac{2i\lambda_{xy} e^{i\phi_{j+1}}}{\sqrt{3} (J_c +4 t_c)}   \hat{a}_{\mu j\uparrow} \hat{a}_{\mu j-1\uparrow}  
		\right] \ket{vac_h}, 
		\end{eqnarray}
		
		In the absence of SMOC  the next manifold of excited states is a sextuplet (a spin triplet with two-fold orbital degeneracy). SMOC splits these states into three doublets with energies $E_4^{3_{ks} E}=t_c + ks\lambda_z/2$, where  $s=-1$, 0, or 1 is the projection of the spin in the $z$-direction (i.e., eigenvalue of $S_z$) in the absence of SMOC and $k$ is the molecular-orbital angular momentum in the absence of SMOC.
		The corresponding wavefunctions are
		\begin{eqnarray}
	\ket{\Phi_4^{3_1E}(0)}	
		&=& \frac{1}{\sqrt{3}} \sum_{j} \left[ 
		e^{i\phi_{j+1}} \hat{a}_{\mu j\downarrow} \hat{a}_{\mu j-1\downarrow} 
		-\frac{\lambda_{xy}}{6 \sqrt{2} t_c}  \left( \hat{a}_{\mu j\uparrow} \hat{a}_{\mu j-1\downarrow} + \hat{a}_{\mu j\downarrow} \hat{a}_{\mu j-1\uparrow}  \right) 	
		\right. \notag\\ && \hspace{1cm} \left.
		-\frac{\lambda_{xy}}{\sqrt{6} (4J_c -2 t_c)}  \left( \hat{a}_{\mu j\uparrow} \hat{a}_{\mu j-1\downarrow} - \hat{a}_{\mu j\downarrow} \hat{a}_{\mu j-1\uparrow}  \right) 
		\right] \ket{vac_h},
%		\\
\end{eqnarray}
\begin{eqnarray}
	\ket{\tilde\Phi_4^{3_1E}(0)}	&=& \frac{1}{\sqrt{3}}\sum_{j} \left[ 
		e^{-i\phi_{j+1}} \hat{a}_{\mu j\uparrow} \hat{a}_{\mu j-1\uparrow} 
		-\frac{\lambda_{xy}}{6\sqrt{2} t_c} \left( \hat{a}_{\mu j\uparrow} \hat{a}_{\mu j-1\downarrow} + \hat{a}_{\mu j\downarrow} \hat{a}_{\mu j-1\uparrow}  \right)
		\right. \notag\\ && \hspace{1cm} \left.
		-\frac{\lambda_{xy}}{\sqrt{6} (4J_c -2 t_c)}  \left( \hat{a}_{\mu j\uparrow} \hat{a}_{\mu j-1\downarrow} - \hat{a}_{\mu j\downarrow} \hat{a}_{\mu j-1\uparrow}  \right) 
		\right] \ket{vac_h}, 	
		\\
%	\end{eqnarray}
%	\begin{eqnarray}	
		\ket{\Phi_4^{3_0E}(-1)}	&=& \frac{1}{\sqrt{6}}\sum_{j} \left[ 
		e^{i\phi_{j+1}} \left( \hat{a}_{\mu j\uparrow} \hat{a}_{\mu j-1\downarrow} + \hat{a}_{\mu j\downarrow} \hat{a}_{\mu j-1\uparrow}  \right) 
		-\frac{\lambda_{xy}}{6 t_c} \hat{a}_{\mu j\uparrow} \hat{a}_{\mu j-1\uparrow}
		\right. \notag\\ && \hspace{1cm} \left.
		+\frac{i\lambda_z e^{i\phi_{j+1}}}{\sqrt{3} (J_c +4 t_c)} \left( \hat{a}_{\mu j\uparrow} \hat{a}_{\mu j-1\downarrow} - \hat{a}_{\mu j\downarrow} \hat{a}_{\mu j-1\uparrow}  \right) 
		\right] \ket{vac_h}, 
		\\
		\ket{\Phi_4^{3_0E}(1)}	&=& \frac{1}{\sqrt{6}}\sum_{j} \left[ 
		e^{-i\phi_{j+1}} \left( \hat{a}_{\mu j\uparrow} \hat{a}_{\mu j-1\downarrow} + \hat{a}_{\mu j\downarrow} \hat{a}_{\mu j-1\uparrow}  \right)
		+\frac{\lambda_{xy}}{6 t_c} \hat{a}_{\mu j\downarrow} \hat{a}_{\mu j-1\downarrow}
		\right. \notag\\ && \hspace{1cm} \left.
		+  \frac{i\lambda_z e^{-i\phi_{j+1}}}{\sqrt{3} (J_c +4 t_c)} \left( \hat{a}_{\mu j\uparrow} \hat{a}_{\mu j-1\downarrow} - \hat{a}_{\mu j\downarrow} \hat{a}_{\mu j-1\uparrow}  \right) 
		\right] \ket{vac_h}, 
		\\	
	\ket{\Phi_4^{3_{-1}E}(1)}	&=& \frac{1}{\sqrt{3}}\sum_{j} \left[ 
	e^{i\phi_{j+1}} \hat{a}_{\mu j\uparrow} \hat{a}_{\mu j-1\uparrow} 
	-\frac{2i\lambda_{xy} e^{-i\phi_{j+1}}}{\sqrt{6} (J_c +4 t_c)}  \left( \hat{a}_{\mu j\uparrow} \hat{a}_{\mu j-1\downarrow} - \hat{a}_{\mu j\downarrow} \hat{a}_{\mu j-1\uparrow}  \right)\right] \ket{vac_h}, 
		\\
	\ket{\Phi_4^{3_{-1}E}(-1)}	&=& \frac{1}{\sqrt{3}}\sum_{j} \left[ 
		e^{-i\phi_{j+1}} \hat{a}_{\mu j\downarrow} \hat{a}_{\mu j-1\downarrow}  
		+\frac{2i\lambda_{xy} e^{i\phi_{j+1}}}{\sqrt{6} (J_c +4 t_c)}  \left( \hat{a}_{\mu j\uparrow} \hat{a}_{\mu j-1\downarrow} - \hat{a}_{\mu j\downarrow} \hat{a}_{\mu j-1\uparrow}  \right)  
		\right] \ket{vac_h}. 
		\end{eqnarray}	
		Note that $\ket{\tilde\Phi_4^{3_1E}(0)}={\mathcal T}\ket{\Phi_4^{3_1E}(0)}$ where ${\mathcal T}$ is the time reversal operator.
		
		Finally, the highest lying excited state with two holes is a singlet with energy 
		$E_4^{1A}=-2 J_c + 2 t_c$ and wavefunction
		\begin{eqnarray}
		\ket{\Phi_4^{1A}(0)}	&=& \frac{1}{\sqrt{6}}\sum_{j} \left[ 
		\left( \hat{a}_{\mu j\uparrow} \hat{a}_{\mu j-1\downarrow} - \hat{a}_{\mu j\downarrow} \hat{a}_{\mu j-1\uparrow}  \right)  
		+ \frac{\lambda}{\sqrt{3} (2 J_c - 4 t_c)}  \left( \hat{a}_{\mu j\uparrow} \hat{a}_{\mu j-1\downarrow} + \hat{a}_{\mu j\downarrow} \hat{a}_{\mu j-1\uparrow}  \right)
		+ \frac{\lambda_{xy} e^{i\phi_{j+1}}}{\sqrt{3} (4 J_c - 2 t_c)}   \hat{a}_{\mu j\downarrow} \hat{a}_{\mu j-1\downarrow} 
		\right. \notag\\&& \hspace{1cm} \left.
		+ \frac{\lambda_{xy} e^{-i\phi_{j+1}}}{\sqrt{3} (4 J_c - 2 t_c)} \hat{a}_{\mu j\uparrow} \hat{a}_{\mu j-1\uparrow} 
		\right] \ket{vac_h}
		\end{eqnarray}	
%	\end{subequations}

In the physically relevant parameter regime, $0<2J_c<t_c$,  we have $E_4^{3A}<E_4^{1E}<E_4^{3E}<E_4^{1A}$.
%\end{widetext}

\section{Cation}\label{append3e}

As we will be interested in the physics due to superexchange between neighboring molecules we also need to understand the physics of the single anion and cation.

For three holes (half-filling) the three-site \tJ model reduces to the  Heisenberg model. 
SMOC does not change the three holes states as  $H_{SMO}^{(\mu)}$ involves moving electrons between Wannier orbitals [cf. Eq. (\ref{HSOreal})] and thus would produce empty sites. Therefore, we are left with the straightforward problem of solving the three-site Heisenberg model.
The eigenstates include two pairs of spin doublets with E symmetry, energy  $E_3^{2E}=-3 J_c/2$ and wavefunctions 
	\begin{eqnarray}
	\ket{\Phi_3^{2E}(k,\sigma)}
	=\frac{1}{\sqrt{3}}\sum_{j=1}^3 e^{i\phi_j k} \hat a_{\mu j\sigma} \hat a_{\mu j\overline\sigma}^\dag \hat a_{\mu 1\overline\sigma}\hat a_{\mu 2\overline\sigma}\hat a_{\mu 3\overline\sigma}\ket{\textrm{vac}_h}, \notag\\
	\end{eqnarray}
	which describes a spin wave  with angular momentum $k=\pm1$ and net spin $\sigma=\pm1/2$.
	The remaining states comprise a fully polarized spin-quadruplet with A orbital symmetry, energy  $E_3^{4A}=0$ and wavefunctions  
	\begin{eqnarray}
	\ket{\Phi_3^{4A}(3/2)}&=&\hat a_{\mu 1\downarrow}\hat a_{\mu 2\downarrow}\hat a_{\mu 3\downarrow}\ket{\textrm{vac}_h},\\
	\ket{\Phi_3^{4A}(1/2)}&=&\frac{1}{\sqrt{3}}\sum_{j=1}^3 \hat a_{\mu j\uparrow} \hat a_{\mu j\downarrow}^\dag \hat a_{\mu 1\downarrow}\hat a_{\mu 2\downarrow}\hat a_{\mu 3\downarrow}\ket{\textrm{vac}_h}  ,\\
	\ket{\Phi_3^{4A}(-1/2)}&=&\frac{1}{\sqrt{3}}\sum_{j=1}^3 \hat a_{\mu j\downarrow} \hat a_{\mu j\uparrow}^\dag \hat a_{\mu 1\uparrow}\hat a_{\mu 2\uparrow}\hat a_{\mu 3\uparrow}\ket{\textrm{vac}_h},\hspace{20pt}
	\end{eqnarray}
	and 
	\begin{eqnarray}
	\ket{\Phi_3^{4A}(-3/2)}&=&\hat a_{\mu 1\uparrow}\hat a_{\mu 2\uparrow}\hat a_{\mu 3\uparrow}\ket{\textrm{vac}_h}. 
	\end{eqnarray}

\section{Anion}\label{append5e}

For one hole the \tJ model reduces to the tight-binding model.
The eigenstates of $H_{tJ}$ can be written as either the Bloch or Condon-Shortley basis states (as these only differ by phase factors) and have energy $\epsilon_k=2t_c\cos{k}$ as expected. 
Clearly, even with SMOC, this problem is straightforward to solve exactly. But,  consistency requires the first order energies and wavefunctions, which are $E_5^{2A}(0,\sigma)=2t_c$, $E_5^{2E}(k,\sigma)=-t_c-k\sigma\lambda_z$, 
%\begin{widetext}
%\begin{subequations}
	\begin{eqnarray}
	\ket{\Phi_5^{2A}(0,\sigma)}&=&
	\frac{1}{\sqrt{3}}\sum_{j=1}^3\left( \hat{a}_{\mu j\overline{\sigma}} +\frac{\sqrt{2}\sigma\lambda_{xy}}{3t_c} e^{i2\sigma\phi_j} \hat{a}_{\mu j{\sigma}} 
	\right)
	%\notag\\&& \hspace*{3.8cm}\times
	\ket{\textrm{vac}_h},
	\end{eqnarray}
	and
	\begin{eqnarray}
	\ket{\Phi_5^{2E}(k,\sigma)}&=&-\frac{1}{\sqrt{3}}\sum_{j=1}^3 \left(k e^{i\phi_j k} \hat{a}_{\mu j\overline{\sigma}} -\frac{\lambda_{xy}\delta_{k,2\overline\sigma}}{3\sqrt{2}t_c} \hat{a}_{\mu j{\sigma}} \right) 
%	\notag\\&& \hspace*{3.8cm}\times
\ket{\textrm{vac}_h},
	\end{eqnarray}
%\end{subequations}
where $k=\pm1$ and $\sigma=\pm1/2$.

\section{Parameters in Eqs. (\ref{Dgen}) }\label{App:Xi}

\begin{eqnarray}
	\Upsilon_{ij} &=& \frac{4}{81(J_c - 2 t_c)t_c}\sum_{g,h=1}^{3}  t_{gh} \left\{ 3t_c (1-\delta_{ig}\delta_{jh})(2\delta_{ig} + 2\delta_{jh} -1) -J_c [\cos(\phi_i-\phi_g) + \cos(\phi_j-\phi_h)]  \right\},
\\
    \Xi^x &=&	  \frac{\lambda _{{xy}}}{81 \sqrt{6} (J_c - 2 t_c)^2 } \Bigg(\left(\frac{J_c}{t_c}\right)^2\Big[-t_{1,2}^2+(5 t_{2,2}-2 t_{2,3}+4 t_{3,1}+3 t_{3,2}+2
    t_{3,3}) t_{1,2}+t_{1,3}^2+t_{2,1}^2+2 t_{2,3}^2-t_{3,1}^2
	 \notag\\&&\hspace*{3cm}
    -2 t_{3,2}^2-4 t_{1,3} t_{2,1} -2 t_{1,3}
    t_{2,2}-5 t_{2,1} t_{2,2}-3 t_{1,3} t_{2,3}-3 t_{2,1} t_{2,3}-4 t_{2,2} t_{2,3}+2 t_{2,2} t_{3,1}
 \notag\\&&\hspace*{3cm}
    -2
    t_{2,3} t_{3,1}+2 t_{1,3} t_{3,2}+2 t_{2,1} t_{3,2}+4 t_{2,2} t_{3,2}+3 t_{3,1} t_{3,2}
    \notag\\&&\hspace*{3cm}
	 +t_{1,1}
	 (-t_{1,2}+t_{1,3}+t_{2,1}-4 t_{2,3}-t_{3,1}+4 t_{3,2})
%	\notag\\&&\hspace*{3cm}
    +(-5 t_{1,3}-2 t_{2,1}-4 t_{2,3}+5
    t_{3,1}+4 t_{3,2}) t_{3,3}\Big]    
\notag\\&&\hspace*{1cm}
    +2 \left(\frac{J_c}{t_c}\right)\Big[-2 t_{1,2}^2-2 (10 t_{2,2}-t_{2,3}+2
    t_{3,1}+3 t_{3,2}+t_{3,3}) t_{1,2}
	 \notag\\&&\hspace*{1cm}
    +2 (t_{1,3}+t_{2,1}){}^2+4 t_{2,3}^2-2 t_{3,1}^2-4
    t_{3,2}^2+2 t_{1,3} t_{2,2}+20 t_{2,1} t_{2,2}+6 t_{1,3} t_{2,3}+6 t_{2,1} t_{2,3}+13 t_{2,2} t_{2,3}
	 \notag\\&&\hspace*{1cm}    
    -2
    t_{2,2} t_{3,1}+2 t_{2,3} t_{3,1}+t_{1,1} (7 t_{1,2}-7 t_{1,3}-7 t_{2,1}+4 t_{2,3}+7 t_{3,1}-4
    t_{3,2})-2 t_{1,3} t_{3,2}-2 t_{2,1} t_{3,2}
	 \notag\\&&\hspace*{1cm}    
    -13 t_{2,2} t_{3,2}-6 t_{3,1} t_{3,2}+(20
    t_{1,3}+2 t_{2,1}+13 t_{2,3}-20 t_{3,1}-13 t_{3,2}) t_{3,3}\Big] 
\notag\\&&\hspace*{1cm}    
    +24  \Big[-t_{2,1}
    t_{2,2}-2 t_{2,3} t_{2,2}-t_{3,1} t_{2,2}+2 t_{3,2} t_{2,2}-3 t_{2,1} t_{2,3}+t_{2,3} t_{3,1}
	 \notag\\&&\hspace*{1cm}    
    +t_{1,1}
    (t_{1,2}-t_{1,3}-t_{2,1}+2 t_{2,3}+t_{3,1}-2 t_{3,2})-t_{2,1} t_{3,2}+3 t_{3,1}
    t_{3,2}+t_{1,3} (2 t_{2,1}+t_{2,2}-3 t_{2,3}-t_{3,2}-t_{3,3})
	 \notag\\&&\hspace*{1cm}    
    +t_{1,2} (t_{2,2}+t_{2,3}-2
    t_{3,1}+3 t_{3,2}-t_{3,3})+(t_{2,1}-2 t_{2,3}+t_{3,1}+2 t_{3,2}) t_{3,3}\Big]
    \Bigg)
      \end{eqnarray}
   \begin{eqnarray} 
   % \\
    \Xi^y &=& 
    \frac{\lambda _{{xy}}}{81\sqrt{2} (J_c - 2 t_c)^2}  \Bigg(\left(\frac{J_c}{t_c}\right)^2\Big[-t_{1,2}^2+(2 t_{1,3}+t_{2,2}+2 t_{2,3}+t_{3,2}+2
    t_{3,3}) t_{1,2}-t_{1,3}^2+t_{2,1}^2+t_{3,1}^2+2 t_{1,3} t_{2,2}
    \notag\\&&\hspace*{3cm}
    -t_{2,1} t_{2,2}+t_{1,3} t_{2,3} 	
    -t_{2,1} t_{2,3}
    -2 t_{2,2} t_{2,3}+3 t_{1,1} (t_{1,2}+t_{1,3}-t_{2,1}-t_{3,1})-2
    t_{2,1} t_{3,1}-2 t_{2,2} t_{3,1}
    \notag\\&&\hspace*{3cm}    	
    -2 t_{2,3} t_{3,1}+2 t_{1,3} t_{3,2}
    -2 t_{2,1} t_{3,2}+2 t_{2,2}
    t_{3,2}-t_{3,1} t_{3,2}+(t_{1,3} 
    -2 t_{2,1}+2 t_{2,3}-t_{3,1}-2 t_{3,2}) t_{3,3}\Big]
    \notag\\&&\hspace*{1cm}     	
    -2 \left(\frac{J_c}{t_c}\right) \Big[2 t_{1,2}^2+2 (2 t_{1,3}+t_{2,2}+t_{2,3}+t_{3,2}+t_{3,3}) t_{1,2}+2 t_{1,3}^2
    -2t_{2,1}^2-2 t_{3,1}^2  +2 t_{1,3} t_{2,2}-2 t_{2,1} t_{2,2}
    \notag\\&&\hspace*{3cm}    	
    +2 t_{1,3} t_{2,3}-2 t_{2,1} t_{2,3}-9 t_{2,2}t_{2,3}+11 t_{1,1} (t_{1,2}+t_{1,3}-t_{2,1}-t_{3,1})-4 t_{2,1} t_{3,1}-2 t_{2,2} t_{3,1}
    \notag\\&&\hspace*{3cm}    	
    -2
    t_{2,3} t_{3,1}+2 t_{1,3} t_{3,2}-2 t_{2,1} t_{3,2}+9 t_{2,2} t_{3,2}-2 t_{3,1} t_{3,2}+(2
    t_{1,3}-2 t_{2,1}+9 t_{2,3}-2 t_{3,1}-9 t_{3,2}) t_{3,3}\Big] 
    \notag\\&&\hspace*{1cm}     	
    +24 \Big[-t_{1,3}
    t_{2,2}-t_{2,1} t_{2,2}+t_{3,1} t_{2,2}+t_{1,3} t_{2,3}-t_{2,1} t_{2,3}+t_{1,1}
    (t_{1,2}+t_{1,3}-t_{2,1}-t_{3,1})-2 t_{2,1} t_{3,1}+t_{2,3} t_{3,1}
    \notag\\&&\hspace*{3cm}      	
    -t_{1,3} t_{3,2}+t_{2,1}
    t_{3,2}-t_{3,1} t_{3,2}+t_{1,2} (2
    t_{1,3}+t_{2,2}-t_{2,3}+t_{3,2}-t_{3,3})+(t_{1,3}+t_{2,1}-t_{3,1}) t_{3,3}\Big]
    \Bigg)    
   \end{eqnarray} 
\end{widetext}

\end{document}